\documentclass[aps,pre,showpacs,showkeys,amsmath,amsfonts,amssymb,notitlepage,superscriptaddress,nofootinbib]{revtex4-1} 
\usepackage[dvips]{graphicx}
\usepackage[applemac]{inputenc}
\usepackage{bbm}
\usepackage{xcolor}
\usepackage[colorlinks,linkcolor=blue,anchorcolor=blue,citecolor=blue,filecolor=blue,menucolor=blue,urlcolor=blue]{hyperref}
\usepackage{breakurl}

\newcommand{\beq}{\begin{equation}} 
\newcommand{\eeq}{\end{equation}}
\newcommand{\bea}{\begin{eqnarray}} 
\newcommand{\eea}{\end{eqnarray}}

\newcommand{\pd}[2]{\frac{\partial #1}{\partial #2}}

\newcommand{\td}[2]{\frac{d #1}{d #2}}

\newcommand{\me}[1]{\left\langle #1 \right\rangle }

\newcommand{\PhSpint}{ \int_{-\infty}^{+\infty} dp \int_{-\pi}^{\pi} d\theta }
\newcommand{\CASES}[1]{\left\lbrace\begin{split}#1\end{split}\right.}

\begin{document}

\title{Violent relaxation in the Hamiltonian mean field model: \\ II. Non-equilibrium phase diagrams} 

\author{Alessandro Santini}
\email{asantini@sissa.it}
\affiliation{SISSA, via Bonomea 265, I-34136 Trieste, Italy}
\author{Guido Giachetti} 
\email{ggiachet@sissa.it} 
\affiliation{SISSA, via Bonomea 265, I-34136 Trieste, Italy}
\affiliation{Istituto Nazionale di Fisica Nucleare (INFN), Sezione di Trieste, via Valerio 2, I-34127 Trieste, Italy}
\author{Lapo Casetti}
\email{lapo.casetti@unifi.it} 
\affiliation{Dipartimento di Fisica e Astronomia, Universit\`a di Firenze, via G.\ Sansone 1, I-50019 Sesto Fiorentino, Italy}  
\affiliation{Istituto Nazionale di Fisica Nucleare (INFN), Sezione di Firenze, via G.\ Sansone 1, I-50019 Sesto Fiorentino, Italy}  
\affiliation{INAF-Osservatorio Astrofisico di Arcetri, Largo E.\ Fermi 5, I-50125 Firenze, Italy}  

\date{\today} 
 
\begin{abstract} 
A classical long-range-interacting $N$-particle system relaxes to thermal equilibrium on time scales growing with $N$; in the limit $N\to \infty$ such a relaxation time diverges. However, a completely non-collisional relaxation process, known as violent relaxation, takes place on a much shorter time scale independent of $N$ and brings the system towards a non-thermal quasi-stationary state. A finite system will eventually reach thermal equilibrium, while an infinite system will remain trapped in the quasi-stationary state forever. For times smaller than the relaxation time the distribution function of the system obeys the collisionless Boltzmann equation, also known as the Vlasov equation. The Vlasov dynamics is invariant under time reversal so that it does not ``naturally'' describe a relaxational dynamics. However, as time grows the dynamics affects smaller and smaller scales in phase space, so that observables not depending upon small-scale details appear as relaxed after a short time. Herewith we present an approximation scheme able to describe violent relaxation in a one-dimensional toy-model, the Hamiltonian Mean Field (HMF). The approach described here generalizes the one proposed in \cite{Giachetti_2019}, that was limited to ``cold'' initial conditions, to generic initial conditions, allowing us to to predict non-equilibrium phase diagrams that turn out to be in good agreement with those obtained from the numerical integration of the Vlasov equation. 
\end{abstract} 
 
 
\maketitle 

\section{Introduction}
\label{sec:intro}
Long-range interactions are those which decay with a slow enough power law of the distance $r$ between the interacting bodies. In particular, systems with interaction decaying slower than $r^{-d}$, where $d$ is the dimension of space, have non-additive energies (see e.g.\ \cite{prl2015}). Paradigmatic examples of such interactions are the gravitational and the electrostatic one, but also dipolar forces in three dimensions or effective interactions between atoms in an optical cavity mediated by the electromagnetic field \cite{njp2016} are long-ranged. The behavior of long-range systems is peculiar both in equilibrium and non-equilibrium, for additive and non-additive systems as well \cite{Campa-Dauxois-Fanelli-Ruffo,CampaEtAl:physrep,LevinEtAlphysrep:2014,eps2021,Giachetti2021}. Here we shall mainly be concerned with non-equilibrium aspects of non-additive long-range systems. The most striking feature of $N$-degree-of-freedom long-range-interacting systems is that the relaxation time $\tau_{\text{rel}}$ to thermal equilibrium\footnote{When thermal equilibrium is not properly defined, as in the case of three-dimensional self-gravitating systems, $\tau_{\text{rel}}$ is the time scale over which the dynamics loses memory of the initial conditions, thus entailing a growth of the Boltzmann entropy.} grows with $N$ and eventually diverges when $N \to \infty$. Such a behavior is a consequence of the fact that mean-field collective effects become more and more important than binary interactions as $N$ grows \cite{Campa-Dauxois-Fanelli-Ruffo}. A large system with long-range interactions will remain out of equilibrium virtually forever, if its initial state is not the thermal equilibrium one. For times smaller than $\tau_{\text{rel}}$ the one-particle distribution function $f(\mathbf{q},\mathbf{p},t)$, where $\mathbf{q}$ and $\mathbf{p}$ are canonically conjugated coordinates and momenta, obeys the non-collisional Boltzmann equation, also referred to as the Vlasov equation. The latter is time-reversal-invariant, so that one may expect that $f$ has a non-relaxational (i.e., oscillatory) dynamics until the effects of binary interactions set in, driving the system towards thermal equilibrium on a time scale $\tau_{\text{rel}}$. On the contrary, the oscillations of $f$ are typically damped on a much shorter, and $N$-independent, time scale, and the system appears to settle in a quasi-stationary state (QSS) that is typically far from a thermal one (see e.g.\ \cite{njp2016,prerap2015,mnras2018} for examples where the QSSs exhibit strongly non-thermal features). Such process is purely non-collisional and was dubbed ``violent relaxation'' by Lynden-Bell who first attempted a theoretical approach to this phenomenon while trying to explain the luminosity profiles of elliptical galaxies \cite{Lynden-Bell:mnras1967}.  Despite decades of research, the problem of fully understanding the mechanism of violent relaxation and of predicting the relation between initial conditions and QSSs is still open (see e.g.\ \cite{Giachetti_2019,LevinEtAlphysrep:2014,PhysRevResearch.2.023379} and references therein). A similar problem is that of Landau damping in non-collisional plasmas, whose dynamics is also described by the Vlasov equation: there, a perturbation damps out because its energy is transferred from the large-scale modes to the individual particles. Violent relaxation is indeed a sort of Landau damping: however, at variance with the latter, that is usually studied in a linear regime and considering perturbations of a homogeneous background, it occurs in a fully nonlinear regime and in inhomogeneous states \cite{BarreOlivettiYamaguchi:jphysa2011,BarreOlivettiYamaguchi:jstat2010}, thus making a full theoretical treatment of this problem extremely difficult. As in Landau damping, as time proceeds the dynamics affects smaller and smaller scales in phase space, so that any macroscopic observable appears to relax to a stationary value although the distribution function never stops evolving. This solves the apparent paradox of a relaxational dynamics governed by a time-reversal-invariant equation and suggests that coarse graining might be a key step towards an effective theory of violent relaxation. An evolution equation for a coarse-grained distribution function was derived in \cite{PhysRevResearch.2.023379}, by imposing that the coarse graining procedure conserves the symplectic structure of phase space. In the case of one-dimensional systems such an equation can be worked out in full detail and allows to make predictions on the scaling of damping times with the coarse graining scale that are in very good agreement with numerical results for a variety of one-dimensional models. 

However, directly solving the evolution equation for the coarse-grained distribution function derived in \cite{PhysRevResearch.2.023379} appears (at least) as complicated as solving the Vlasov equation itself, so that devising less general but easier to solve approximation schemes able to gain some insight into the violent relaxation process is still very useful. An example of an approximation scheme based on introducing suitable moments of the distribution function and then realizing the coarse graining by truncating the hierarchy of moments at a given level and introducing an effective dissipation was proposed in \cite{Giachetti_2019} (from now on referred to as Paper I) and applied to the Hamiltonian Mean Field (HMF) model, one of the most studied examples of systems with long-range interactions. There only ``cold collapse'' was considered, i.e., the dynamics resulting from initial conditions with zero kinetic energy; the aim of the present paper is to extend the approach of Paper I to generic initial conditions, so that non-equilibrium phase diagrams depending on the choice of the initial condition can be worked out and compared with the outcomes of the numerical integration of the Vlasov equation. Actually we shall derive more than simple phase diagrams: we shall calculate the values of the order parameter (the magnetization, see Sec.\ \ref{sec:HMF} for the details) in the QSS and plot them in a plane where each point corresponds to a given initial state using a color code such that black (resp.\ colorful) corresponds to zero (resp.\ nonzero) order parameter: phase boundaries will then correspond to boundaries within colored and black regions in the diagram. 

The paper is structured as follows: in Sec.\ \ref{sec:HMF} the HMF model is described, in Sec.\ \ref{sec:moments} we briefly recall the theoretical approach introduced in Paper I, in Sec.\ \ref{sec:phasediag} we derive phase diagrams by explicitly implementing the approximation scheme at the leading order and at the next-to-leading order for some classes of initial conditions, in Sec.\ \ref{sec:comparison} we compare our theoretical results with the outcomes of numerical simulations of the Vlasov equation, and finally in Sec.\ \ref{sec:concl} we draw our conclusions and discuss open issues and possible developments.

\section{Hamiltonian mean field model}
\label{sec:HMF} 

The Hamiltonian Mean Field (HMF) model is a toy model that has become a cornerstone in the study of long-range-interacting systems, due to its simplicity together with the richness of its dynamics. According to Chavanis and Campa \cite{ChavanisCampa} the model was firstly introduced by Messer and Spohn \cite{MesserSpohn}, who called it the ``cosine model''. The model was actually made popular and the name and HMF acronym introduced after more than ten years by Antoni and Ruffo in a seminal work \cite{RUFFO1995}.

The HMF model consists in a fully connected network of $N$ classical planar rotators and is defined by the following Hamiltonian:
\begin{equation}
\mathcal{H} = \sum_{i=1}^N \frac{p_i^2}{2} -\frac{J}{2N}\sum_{i,j=1}^N \cos(\theta_i-\theta_j)\,, \label{eq:hmf_hamiltonian}
\end{equation}
where $\theta_i \in [-\pi,\pi)$ are angular coordinates and $p_i$ are their conjugated momenta, for $i=1,\ldots,N$; for simplicity we have set the rotational inertia moment of each rotator equal to unity. The coupling constant $J$ is divided by $N$ to ensure the extensivity of the energy (Kac rescaling, see e.g.\ \cite{Campa-Dauxois-Fanelli-Ruffo}). The model can be also seen as either representing $N$ particles confined on a ring of unit radius and interacting via a cosine potential, or a classical fully connected XY model both in the attractive $J>0$ ferromagnetic case and in the repulsive $J<0$ anti-ferromagnetic case. Indeed, the Hamiltonian \eqref{eq:hmf_hamiltonian} can be written as 
\begin{align}
&\mathcal{H}=\sum_{i=1}^N \frac{p_i^2}{2} -\frac{J}{2N} \sum_{i,j=1}^N \mathbf{S}_i\cdot\mathbf{S}_j\,,
\end{align}
where the $\mathbf{S}_i=\left(S_i^{(x)},S_i^{(y)}\right)$ are planar spins with unit norm $\mathbf{S}_i^2=1$, for all $i=1,\ldots,N$.
The Hamiltonian is invariant under the $O(2)$ continuous symmetry group.  In what follows we will be interested only in the ferromagnetic case so that $J>0$, and we shall set $J = 1$ in order to fix the energy scale to unit. In thermal equilibrium, the ferromagnetic HMF model exhibits a phase transition with spontaneous breaking of the $O(2)$ symmetry at $T_c = 1/2$. The order parameter is the magnetization per spin
\begin{align}
&\mathbf{m}=\frac{1}{N}\sum_{i=1}^N\mathbf{S}_i=(m_x,\ m_y), \end{align}
where\begin{subequations}\begin{align}
&m_x=\frac{1}{N}\sum_{i=1}^N\cos\theta_i = \me{\cos\theta},\\
&m_y=\frac{1}{N}\sum_{i=1}^N\sin\theta_i = \me{\sin\theta}.
\end{align}\label{eqs:magnetisation_components}\end{subequations}
The trigonometric identity for the cosine $\cos(\theta_i-\theta_j)=\cos\theta_i\cos\theta_j+\sin\theta_i\sin\theta_j$ allows us to express the interaction term of the Hamiltonian \eqref{eq:hmf_hamiltonian} only in terms of the order parameter
\begin{align}
\mathcal{H} =\sum_{i=1}^N\frac{p_i^2}{2}-\frac{1}{2}N\mathbf{m}^2.
\end{align}
The equations of motion of the system are given by
\begin{subequations}\begin{align}
&\dot{\theta}_i = p_i \\
&\dot{p}_i = - m_x \sin\theta_i +m_y \cos\theta_i
\end{align}\end{subequations}
where $- m_x \sin\theta_i +m_y \cos\theta_i\equiv F_i$ is the force acting on the $i$-th particle and it is apparent that the interaction among particles comes only from the magnetization, hence the ``mean field'' in the name of the model.

In the limit $N\to\infty$ we can introduce the single-particle distribution function $f(\theta,p,t)$ and replace the averages in equations \eqref{eqs:magnetisation_components} with phase space averages weighted by $f$. The two components of the magnetisation become the following functionals of $f$:
\begin{subequations}\begin{align}
&m_x[f] = \PhSpint f(\theta,p,t) \cos\theta = \me{\cos\theta},\\ 
&m_y[f] = \PhSpint f(\theta,p,t) \sin\theta = \me{\sin\theta}.
\end{align}\end{subequations}
Furthermore, we can define a mean-field potential\begin{equation}
V[f](\theta) = -\int_{-\infty}^\infty dp' \int_{-\pi}^\pi d\theta' f(\theta',p',t) \cos(\theta-\theta') = -\cos \theta \me{\cos \theta} - \sin \theta \me{\sin\theta} = -m_x[f]\cos \theta  - m_y[f]\sin\theta,
\end{equation}
from which we can derive the mean-field force field\begin{equation}
F[f](\theta) = -\td{V[f](\theta)}{\theta} = -m_x[f]\sin\theta +m_y[f]\cos\theta,
\end{equation}
so that the distribution function $f(\theta,p,t)$ evolves in time according to the Vlasov equation\begin{equation}
\pd{f}{t} + p \pd{f}{\theta} + F[f](\theta) \pd{f}{p} = 0.
\label{eq:Vlasov_HMF}
\end{equation}
For a system with large but finite $N$, the Vlasov equation \eqref{eq:Vlasov_HMF} will be valid only for $t < \tau_{\text{rel}}$.  
We will consider initial conditions which are symmetric around the origin of the phase space, such that $f(\theta,p,0)=f(-\theta,-p,0)$. Such invariance is conserved in the evolution of the system, so that the distribution function will be such that\begin{equation}
f(\theta,p,t)=f(-\theta,-p,t)\label{eq:symmetry_assumption}
\end{equation}
at any time $t$. This implies a vanishing total momentum of the system and $m_y \equiv 0$, so that $\mathbf{m}$ will always be along the $\theta=0$ axis and $m\equiv m_x$. The Vlasov equation will then become\begin{align}
    \pd{f}{t}+p\pd{f}{\theta}-m[f]\sin\theta \pd{f}{p}=0
    \label{eq:VlasovHMF}
\end{align}
with the boundary condition $f(\theta,p,t)=f(\theta+2\pi,p,t)$.

\section{Coarse graining the distribution function via its moments of inertia}
\label{sec:moments} 

Let us now recall the approximation scheme introduced in Paper I to deal with the cold collapse case, that we shall later generalize to consider generic initial conditions. We refer the reader to Paper I for further details. 

We define the generalized moments of inertia $I_{k,n}(t)$ of the distribution function as \begin{equation}
    I_{k,n}(t) = \PhSpint f(\theta,p,t) \theta^k p^n = \me{\theta^k p^n}\,; \label{eq:moments_def}
\end{equation}
our symmetry assumption \eqref{eq:symmetry_assumption} implies that $I_{k,n}(t) = 0$ if $k+n$ is odd. High-order moments, i.e., $\me{\theta^kp^n}$ with $k+n\gg 1$, describe the finer details of the distribution function; low-order moments describe large-scale, macroscopic features. 
Among the low-order moments we find: $I_{0,0}$, which is the norm of $f$ (equal to $1$ at any time $t$); $I_{0,2}=\me{p^2}$, which is proportional to the kinetic energy; $I_{2,0}=\me{\theta^2}$, which measures the width in $\theta$ of $f$; $I_{1,1}=\me{\theta p}$, the covariance of positions and velocities. Since the dynamics moves to smaller and smaller scales as time proceeds, we expect these low-order moments to settle down to a stationary value before the higher-order ones.

We now want to find an evolution equation for the inertia moments. To do so, first of all we replace the $\sin\theta$ term in the force field of the Vlasov equation \eqref{eq:VlasovHMF} with its Taylor expansion up to a finite order $2J+1$, obtaining
\begin{equation}
    \pd{f}{t} + p\pd{f}{\theta} - m \left[\sum_{j=0}^{J} \frac{(-1)^j}{(2j+1)!}\theta^{2j+1}\right] \pd{f}{p} = 0. \label{eq:taylor}
\end{equation}
We note that the leading order $J=0$ is equivalent to a harmonic approximation, i.e., $V[f](\theta) = -m\cos\theta \approx -m\left(1-\theta^2/2\right)$. 
As shown in Paper I, using Eq.\ \eqref{eq:taylor} and the definition \eqref{eq:moments_def} of the moments  $I_{k,n}$  we get the evolution equations of the latter as
\begin{equation}
    \dot{I}_{k,n} = k I_{k-1,n+1} -nm\sum_{j=0}^J\frac{(-1)^j}{(2j+1)!}I_{k+2j+1,n-1},\label{eq:evo-moments}
\end{equation}
where the magnetisation $m$ is given by\begin{equation}
    m = \sum_{j=0}^{J+1} \frac{(-1)^j}{(2j)!}I_{2j,0}.\label{eq:m-expansion}
\end{equation}
Equations \eqref{eq:evo-moments} and \eqref{eq:m-expansion} delineate the hierarchy in the interactions among different moments. Indeed, from Eq.\ \eqref{eq:evo-moments}, it is apparent that a moment of a certain order $J$ strongly interacts with ``nearest-neighbour'' moments while it weakly interacts with higher order ones. However, low-order moments play a special role in this picture since they drive the evolution of the magnetisation, being
\begin{equation}
    m=1-\frac{1}{2}I_{2,0}+\frac{1}{24}I_{4,0}-\frac{1}{720}I_{6,0}+\frac{1}{40320}I_{8,0} + \cdots\, ,
\end{equation}
and in doing so they strongly interact, drive and force the evolution of all the moments. Fine details of the distribution function are not important to determine the value of a macroscopic observable. Therefore we can perform a coarse graining on the dynamics by truncating the hierarchy of moments at a given order and neglecting all the higher-order ones. Virial macroscopic oscillations are described by the contribution of a few low-order moments, that interact strongly with each other. Higher-order moments are forced by the low-order ones via their coupling with the magnetization $m$, but their backreaction on the low-order moments is weak and can be taken into account as an effective dissipation. 

The evolution equations \eqref{eq:evo-moments} are a finite set of equations invariant under time reversal, so that they cannot show dissipation. The simplest way to model the effective dissipation due to all the higher-order terms is to add a damping therm to the evolution of $p$-odd moments, i.e., replace Eqs.\ \eqref{eq:evo-moments} with 
\begin{subequations}\begin{align}
\dot{I}_{k,n} &= k I_{k-1,n+1} -nm\sum_{j=0}^J\frac{(-1)^j}{(2j+1)!}I_{k+2j+1,n-1} - \gamma_{k+n} I_{k,n}\,; \\
\gamma_{k+n} &= 0\ \text{if $n$ is even}.
\end{align}\label{eq:evo-diss-I_kn}
\end{subequations}
In this picture, the $\gamma_{k+n}$ cannot be derived from first principles so that they are parameters of the theoretical model and are expected to be different for moments of different order, since the dissipation time scale of higher-order moments is expected to be longer than that of lower-order moments. We note that with this choice of friction coefficients the energy density $\varepsilon$ is conserved at any truncation order $J$, i.e.,
\begin{equation}
\td{\varepsilon}{t} = \td{}{t}\left(\frac{1}{2} \me{p^2} -\frac{1}{2} m^2\right) = \frac{1}{2} \dot{I}_{0,2} - m \dot{m} = 0\, ;\label{cons energy}
\end{equation}
this immediately follows from equations \eqref{eq:evo-diss-I_kn} and \eqref{eq:m-expansion}, because
\begin{subequations}
\begin{align}
\dot{I}_{0,2} &= -2m \sum_{j=0}^J \frac{(-1)^j}{(2j+1)!}I_{2j+1,1},\\
\dot{m} &= \sum_{j=0}^{J+1} \frac{(-1)^j}{(2j)!}\dot{I}_{2j,0}=\sum_{j=0}^{J+1} \frac{(-1)^j}{(2j)!} 2j I_{2j-1,1} = \sum_{j=0}^J \frac{(-1)^{j+1}}{(2j+1)!}I_{2j+1,1} = \frac{\dot{I}_{0,2}}{2m}\, .
\end{align}\label{eqs:cons_energy}
\end{subequations}
As worked out in detail in Paper I, the effective dissipation, at least at the lowest order, can be shown to arise from the evolution equations of the moments under reasonable assumptions on the dynamics of the higher-order moments. In close analogy with the Caldeira-Legget mechanism, this involves a renormalization of the lowest-order dynamics as well. However, since we want to focus on the extension of the method to generic initial conditions, we will stick to the simplest implementation of the effective dissipation given by Eqs.\ \eqref{eq:evo-diss-I_kn}. This will also allow us to consider the next-to-leading order without changing the evolution equations. Before going on, we note that, as mentioned in the Introduction, in \cite{PhysRevResearch.2.023379} a much more general coarse graining procedure is presented, deriving a general equation for the evolution of a coarse-grained distribution function, which is however not analytically tractable. The procedure presented in Paper I and generalized here can be seen as an approximate and more easily tractable realization of the general scheme discussed in \cite{PhysRevResearch.2.023379}, although it is not easy to derive it directly from the general evolution equation. 

\section{non-equilibrium phase diagrams of the HMF model}
\label{sec:phasediag} 
We shall now use the evolution equations for the moments of the distribution function to derive the QSSs corresponding to given choices of initial conditions, going beyond the case of ``cold collapse'' considered in Paper I. We shall start with the leading-order approximation and then proceed to the next-to-leading-order approximation. 

\subsection{Leading-order approximation}
\label{sec:LO} 
At the leading order, $n+k\leq2$, the only relevant inertia moments are \begin{subequations}
\begin{align}
    &x\equiv I_{2,0} = \me{\theta^2},\\
    &y\equiv I_{1,1} = \me{\theta p},\\
    &z\equiv I_{0,2} = \me{p^2},
\end{align}
\end{subequations}
while the magnetisation is given by\begin{equation}
    m \approx 1-\frac{1}{2} I_{2,0} = 1-\frac{1}{2}x,
\end{equation}
and the evolution equations \eqref{eq:evo-moments} become
\begin{subequations}
\begin{align}
    &\dot{x} = 2y\,,\label{eq1:evo_leading-order}\\
    &\dot{y} = z - \left(1-\frac{1}{2}x\right)x\label{eq2:evo_leading-order}-\gamma_{2}y\,,\\
    &\dot{z} = -2\left(1-\frac{1}{2}x\right)y\,.
\end{align}\label{eqs:evo_leading-order}
\end{subequations}
As shown in \eqref{eqs:cons_energy} the above equations admit the following integral of motion\begin{equation}
    \varepsilon = \frac{1}{2} z -\frac{1}{2}\left(1-\frac{1}{2}x\right)^2\,,\label{eq:energy_leading-order}
\end{equation}
which is the energy density. The conservation of $\varepsilon$ allows us to get rid of a variable in the system \eqref{eqs:evo_leading-order}. Indeed, we can replace $z$ with $2\varepsilon+\left(1-x/2\right)^2$ and substitute it in \eqref{eq2:evo_leading-order}; then, observing that \eqref{eq1:evo_leading-order} implies $\ddot{x}=2\dot{y}$, we can finally recast Eqs.\ \eqref{eqs:evo_leading-order} in the Newtonian form\begin{equation}
    \frac{1}{2}\ddot{x} = -\td{V_\mathrm{eff}(x)}{x}-\gamma_{2} \frac{1}{2}\dot{x}\,,
\end{equation}
where the effective potential $V_\mathrm{eff}(x)$ is given by\begin{equation}
    V_\mathrm{eff}(x) = -\left(2\varepsilon +1\right)x+x^2-\frac{1}{4}x^3 = -xz\,.
\end{equation}
We are thus describing the evolution of the lowest-order inertia moments as the damped motion of a fictitious particle in an effective potential. For a vanishing damping term $\gamma_{2} =0 $ the conserved energy $\Lambda$ associated to this motion would be given by
\begin{equation}
    \Lambda = \frac{1}{4}\dot{x}^2+V_\mathrm{eff}(x)=y^2-xz\leq0 \, .
\end{equation}
The effect of the friction is to dissipate the initial energy $\Lambda$ until the particle eventually sets in a minimum $x_\mathrm{min}$ of the potential $V_\mathrm{eff}(x)$ which corresponds to a quasi-stationary magnetisation $m_\mathrm{qss}=1-x_\mathrm{min}/2$ and is given by the solution of 
\begin{equation}
    -\td{V_\mathrm{eff}}{x} = \frac{3}{4}x^2-2x+(2\varepsilon+1)=0\,,
\end{equation}
so that\begin{equation}
    x_\mathrm{min} = \frac{2}{3}\left(2\pm\sqrt{1-6\varepsilon}\right)\,.
\end{equation}
\begin{center}
\end{center}
\begin{figure}
    \centering
    \includegraphics[width=0.6\linewidth]{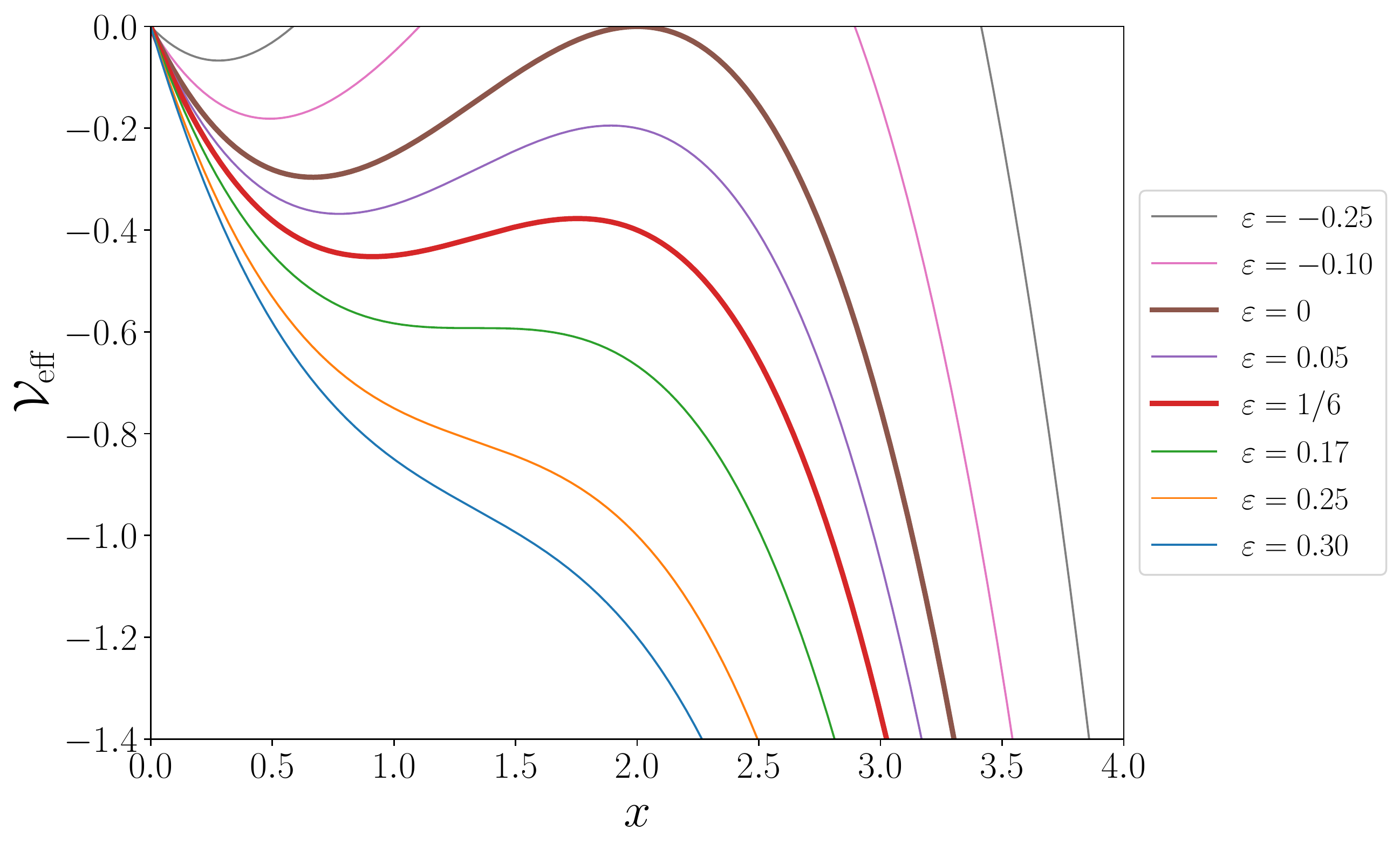}
    \caption{Graph of $V_\mathrm{eff}$ for different values of $\varepsilon$. It is apparent that the motion of the particle cannot be bounded for large enough energy densities. Moreover, since $m=1-x/2$ and since the minimum is in the region $x<2$, we cannot describe quasi-stationary states with negative magnetizations $m<0$. 
    }
    \label{fig:Veff_not-reflected}
\end{figure}
What we have seen until now is consistent if the motion of the fictitious particle is bounded. As long as initial conditions are cold, i.e., with vanishing initial kinetic energy, as considered in Paper I, there are no problems, but being $V_\mathrm{eff}(x)\approx -x^3/4$ for large $x$ this approach breaks down for warm enough initial conditions. Requiring that $x_\mathrm{min}$ is real implies $\varepsilon<1/6$; moreover, the energy density is bounded from below since $\varepsilon=\me{p^2}/2-m^2/2$ implies $\varepsilon>-1/2$. We can thus describe only a slice of the phase diagram for $\varepsilon \in [-1/2,\ 1/6]$. In addition to that, we can only describe the evolution for $m\geq0$ (see Figure \ref{fig:Veff_not-reflected} where the effective potential is plotted against $x$ for different values of $\varepsilon$). 

Therefore, for warm enough initial conditions, i.e., with sufficiently large initial kinetic energy, our approximation breaks down, since in these cases the system could reverse its magnetization. 
We can however work around this problem exploiting a symmetry of the HMF model. Indeed, any state with $m<0$ of the HMF model is perfectly equivalent to a collapsed state with $m>0$ thanks to the following transformation
\begin{subequations}\begin{align}
    \theta' &= \pi - \theta, \\
    p'      &= -p.
\end{align}\label{eq:transformation}
\end{subequations}
This change of variables transforms $\cos\theta$ into $-\cos\theta'$, implying $m'=-m$, while $\sin\theta'=\sin\theta$. Hence this transformation acts on the Vlasov equation such as\begin{equation}
    \pd{f}{t}+p\pd{f}{\theta}-m\sin\theta\pd{f}{p}=0 \longrightarrow \pd{f}{t}+p'\pd{f}{\theta'}-m\sin\theta'\pd{f}{p'}=0
\end{equation}
so that the equation is sactually invariant under such a transformation. This means that any collapsed state with a negative magnetisation, i.e., a state peaked in $\pm \pi$, is equivalent to a collapsed state peaked in the origin by means of the transformation \eqref{eq:transformation}. 
We can thus adopt the following simple prescription: we change the sign of $p$-odd moments, according to the symmetry of the transformations in \eqref{eq:transformation}, each time that $m\to -m$. Indeed, with this prescription we effectively bind the motion of our fictitious particle in the region with $m>0$ and prevent the particle to run away from the physical region.

\begin{figure}
    \centering
    \includegraphics[width=0.6\linewidth]{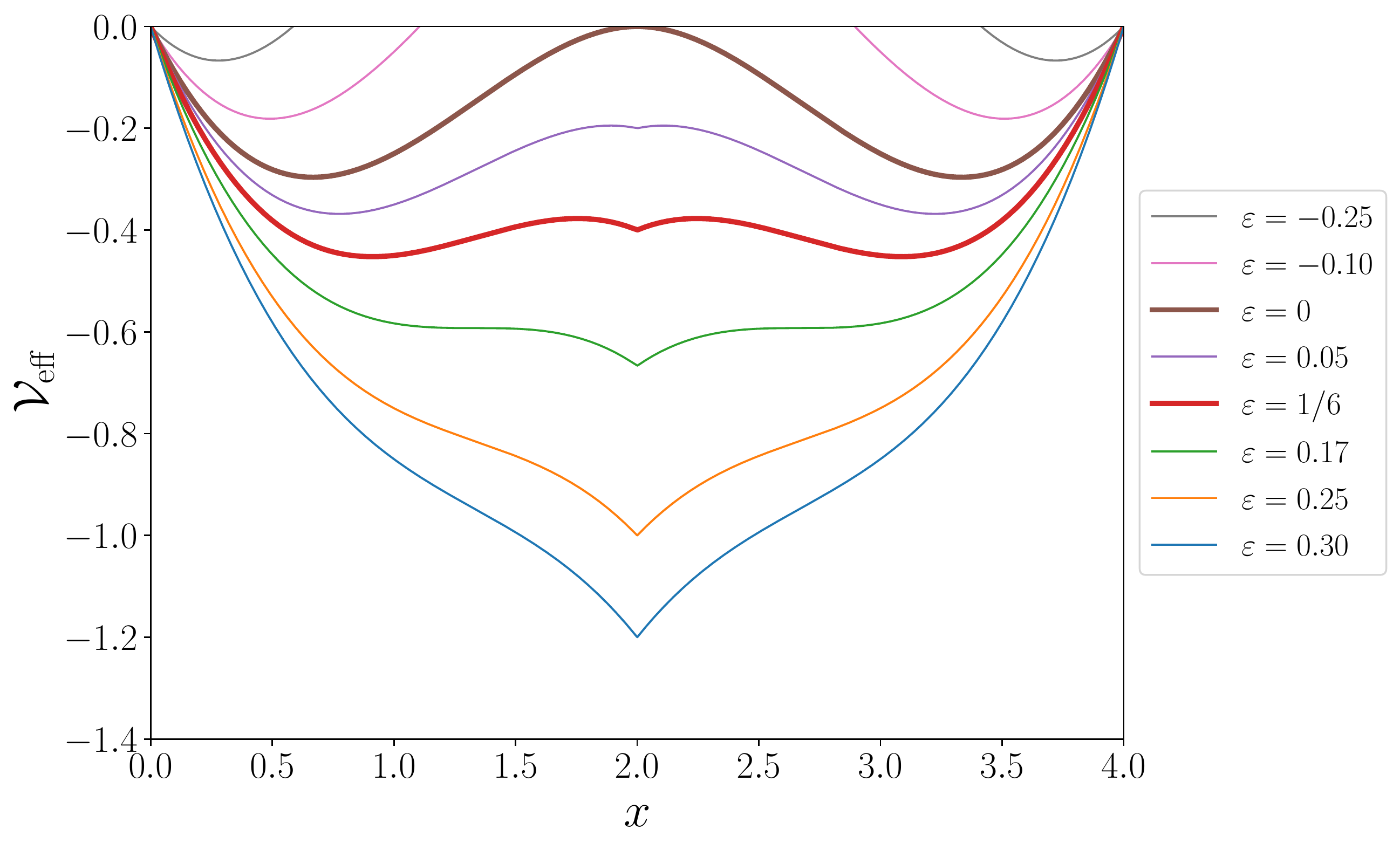}
    \caption{Graph of the modified effective potential $\mathcal{V}_\mathrm{eff}$ for different values of $\varepsilon$ as a function of $x$. When $\varepsilon\leq0$ we are in the cold region in which we find two separate basins of attraction with two symmetric minima. The warm region is found for $0<\varepsilon<1/6$, where there are three minima, one in $x=2$ (corresponding to $m=0$) and the other two symmetric with respect to $x = 0$. Finally for $\varepsilon\geq 1/6$ only the minimum in the non-analytic point $x=2$ of the effective potential remains. 
    }
    \label{fig:Veff_reflected}
\end{figure}

At the leading order only $y=\me{\theta p}$, proportional to the velocity of the fictitious particle, has to change sign when $m$ changes sign (that is, whenever the particle crosses the point $x=2$). This can be accomplished inserting an elastic wall in $x=2$ which confines the motion of the particle in $x\in[0,\ 2]$, or, in terms of the magnetisation, in $m\in[0,\ 1]$. Even better, being the effective system one-dimensional, at this leading order we can take into account also the sign of the magnetisation by reflecting the potential around $x=2$, that is, considering the modified effective potential
\begin{equation}
    \mathcal{V}_\mathrm{eff}(x) = \CASES{&V_\mathrm{eff}(x) && \forall\, x \in [0,2]; \\ &V_\mathrm{eff}(x-4) && \forall\, x \in [2,4].} 
\end{equation}
Moreover, since $\Lambda\leq 0$ the motion of the particle is bounded. The new effective potential $\mathcal{V}_\mathrm{eff}(x)$ is plotted against $x$ in figure \ref{fig:Veff_reflected} for different values of $\varepsilon$.

When $\varepsilon\leq0$ we are in the ``cold region''; here the accessible zone in which the fictitious particle can move is partitioned into two intervals $I_1 \subset [0,2]$ and $I_2 \subset [2,4]$, corresponding to $m>0$ and $m<0$, respectively. The system  cannot reverse its magnetisation and, thanks to the damping, will eventually settle down in one of the two minima, depending on the sign of the initial magnetisation $m_0$. We thus obtain an analytic prediction\footnote{As shown in Paper I, an even better analytic prediction can be obtained at this leading order if we take into account the renormalization of the dynamics obtained by a more refined treatment of the effective dissipation.} of the quasi-stationary magnetisation:
\begin{equation}
m_\mathrm{qss} = \pm \frac{1+\sqrt{1-6\varepsilon}}{3}\,.\label{eq:m_qss-leading-order}
\end{equation}
Therefore when $\varepsilon\leq0$ the system is in a ferromagnetic phase: the magnetization of the quasi-stationary state is nonzero and bounded such that $2/3\leq\left\lvert m_\mathrm{qss}\right\rvert\leq 1$. At the edge of this zone, when $\varepsilon=0$, another equilibrium appears, located in $x=2$ or $m_\mathrm{qss}=0$, but it is unstable so that the damped particle will never be able to reach it if $m_0 \neq 0$. Note that in the cold region the predicted stationary value \eqref{eq:m_qss-leading-order} of the magnetization in the quasi-stationary state only depends on $\varepsilon$.

Increasing the value of the energy density $\varepsilon$ we enter the ``warm region'' where $\varepsilon \in	\left(0, 1/6\right)$. Here the effective potential $\mathcal{V}_{\mathrm{eff}}$ exhibits three minima: one, non-analytic, in correspondence of $x=2$, i.e., $m=0$, and two other ones symmetric with respect to the former. In this region, for some initial conditions, the system reverses its magnetisation: this happens either for large or small values of $m_0$ and for $y=0$, or for large values of the initial correlation $y=\me{\theta p}$, proportional to the velocity $\dot{x}$ of the fictitious particle. In this case the final state of the system is not easily predictable, but we can suppose the probability of a given state to occur to be proportional to the amplitude of its attraction basin. In the warm region both ferromagnetic, where $m_\mathrm{qss}$ is given by Eq.\  \eqref{eq:m_qss-leading-order}, and paramagnetic, i.e., with $m_\mathrm{qss}=0$, phases may occur, depending on the initial conditions.

Eventually, by increasing $\varepsilon$ we reach the ``hot region''  $\varepsilon \geq 1/6$. Here, although the system always reverses its magnetisation $m$ during virial oscillations, there is only one minimum of the potential in $x=2$, which means that the prediction for $m$ in the quasi-stationary state is $m_\mathrm{qss}=0$, so that the only possible phase allowed in this zone is the paramagnetic one. 

\begin{figure}
    \centering
    \includegraphics[width=0.49\linewidth]{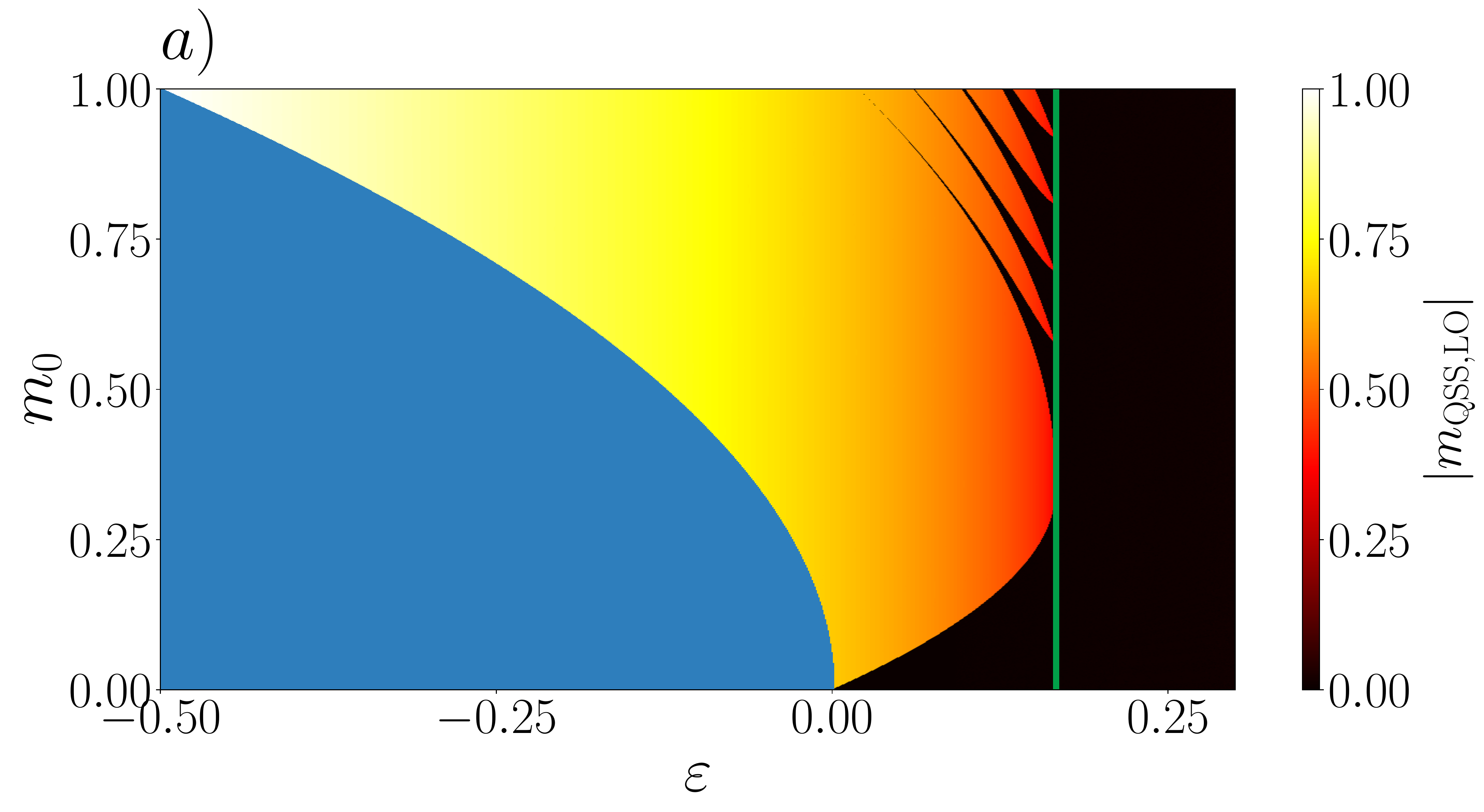}
    \includegraphics[width=0.49\linewidth]{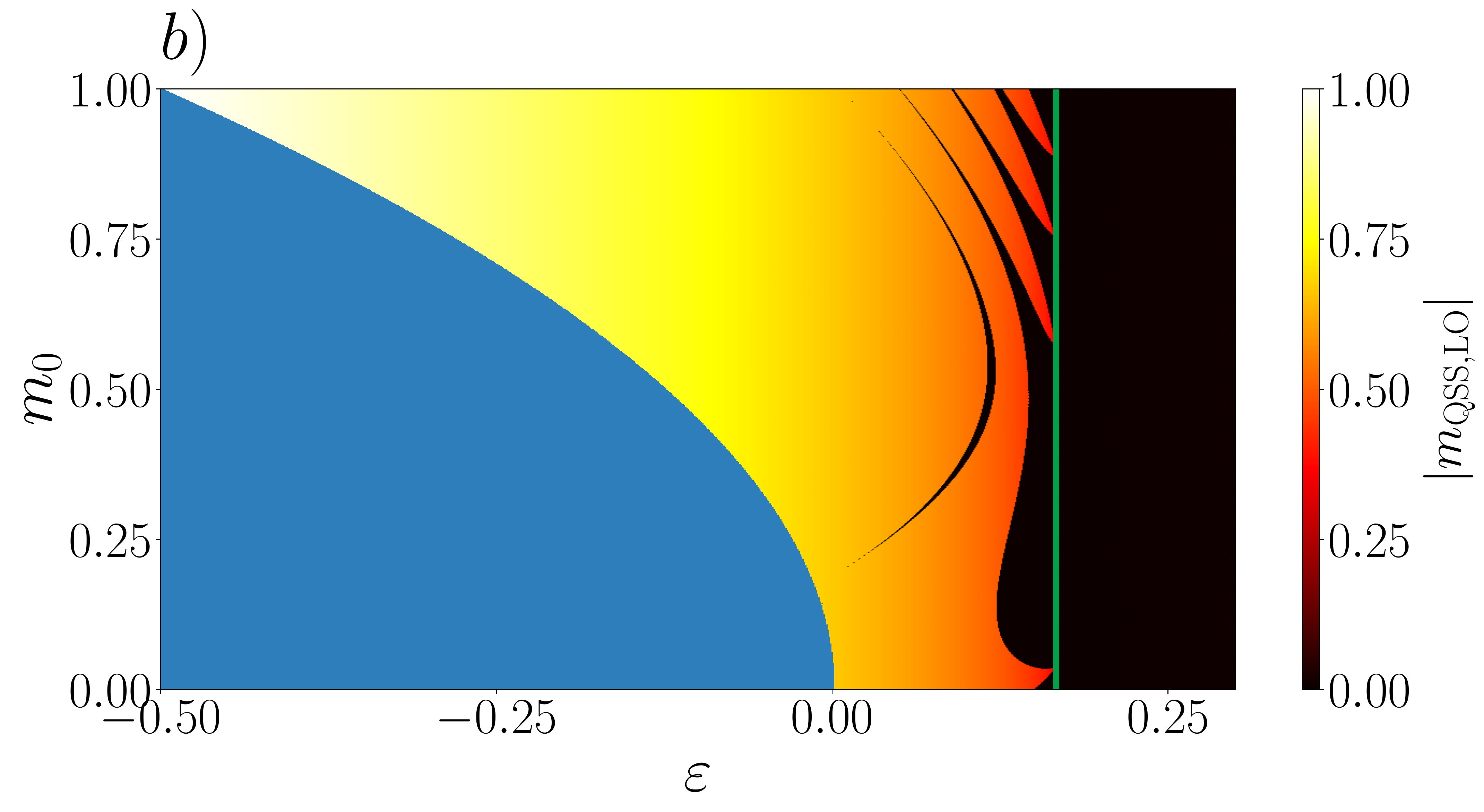}
    \caption{Examples of leading-order predictions for the phase diagrams. In both cases $\gamma_{2} =0.1$; as concerns the initial covariance $y_0=\me{\theta p}_0$, the panel $a)$ was obtained with $y_0=0$ while panel $b)$ with $y_0=0.25$. The green line marks the separation between the hot and the warm regions at $\varepsilon=1/6$ while the blue region of the diagram is the non-physical zone in which the kinetic energy would be negative.}
    \label{fig:Theoretical_pd_lo}
\end{figure}

At the leading order of our approximation we are then able to analytically predict the value of $m$ after violent relaxation to the quasi-stationary state either when $\varepsilon < 0$ (cold region) using Eq.\ \eqref{eq:m_qss-leading-order} or when $\varepsilon > 1/6$ (hot region), where $m_{\text{qss}} = 0$. In both previous cases $m_{\text{qss}}$ depends only on $\varepsilon$. In the warm region $0 < \varepsilon < 1/6$ the value of $m$ after violent relaxation also depends on the initial magnetization $m_0$ or on the initial covariance $y_0=\me{\theta p}_0$, not only on the energy density. To obtain the values of $m_{\text{qss}}$ in the warm region we solved the system of differential equations \eqref{eqs:evo_leading-order} with a standard $4^\mathrm{th}$-order Runge-Kutta algorithm \cite{press2007numerical}. We can thus plot phase diagrams on the $m_0$-$\varepsilon$ plane with a color code to indicate the value of the magnetization, where black corresponds to zero magnetization (paramagnetic quasi-stationary state) and white to the maximum possible magnetization ($m_{\text{qss}} = 1$), so that these phase diagrams contain also the information on the value of the magnetization and not only the location of phase boundaries. In Fig.\ \ref{fig:Theoretical_pd_lo} the predicted leading-order non-equilibrium phase diagram is shown for two classes of initial conditions (see caption for details). 
It is apparent that there is always a sharp boundary between coloured (ferromagnetic) and black (paramagnetic) regions: this means that our approach predicts discontinuous non-equilibrium transitions. The fringes in the phase boundary, that is, the re-entrant tongues in the warm region of phase diagram, are an intriguing prediction of our approximation: as we shall see in the following, these features of the phase diagram survive with little modification also at the next-to-leading order, but do not compare very well to the shape of the phase boundary obtained by integrating the Vlasov dynamics, at least for the initial conditions we investigated. Nonetheless, also the ``true'' Vlasov phase diagrams may exhibit very complicated phase boundaries in the warm region and also regions where the phase boundary exhibit fringes that look qualitatively very similar to the ones predicted by our theoretical approach (see Sec.\ \ref{sec:fringes}).   

\subsection{Next-to-leading-order approximation}
\label{sec:NLO} 

In principle it is possible to write the set of evolution equations for the inertia moments at any given order. The next-to-leading order corresponds to $J=1$ or $n+k\leq 4$. In this case the evolution equations \eqref{eq:evo-diss-I_kn} of the inertia moments become a system of eight coupled differential equations which reads as
\begin{subequations}
\begin{align}
\dot{I}_{2,0} &= 2 I_{1,1}\,,\\
\dot{I}_{1,1} &= I_{0,2} - m \biggl( I_{2,0} -\frac{1}{6} I_{4,0} \biggr)-\gamma_2 I_{1,1}\,, \\
\dot{I}_{0,2} &= -2m\biggl( I_{1,1} -\frac{1}{6} I_{3,1} \biggr)\,, \\
\dot{I}_{4,0} &= 4 I_{3,1}\,, \\ 
\dot{I}_{3,1} &= 3 I_{2,2} - m I_{4,0} -\gamma_{4} I_{3,1}\,, \\
\dot{I}_{2,2} &= 2 I_{1,3} - 2 m I_{3,1}\,, \\
\dot{I}_{1,3} &= I_{0,4} - 3 m I_{2,2} - \gamma_{4} I_{1,3}\,, \\
\dot{I}_{0,4} &= -4 m I_{1,3}\,,
\end{align}\label{eqs:evo_nlo}
\end{subequations}
where the magnetisation $m$ is now given by\begin{equation}
m = 1 - \frac{1}{2} I_{2,0} + \frac{1}{24} I_{4,0}\,.
\end{equation}
In order to describe the correct behaviour of $m$, avoiding the breakdown of our approximation for warm initial conditions, we use the same trick as at the leading order: when $m$ reaches zero we change the sign of the $p$-odd moments, i.e., $I_{1,1}$, $I_{1,3}$ and $I_{3,1}$. Thus, we are again describing the behaviour of $\lvert m(t)\rvert$. Moreover, the following relation holds:
\begin{equation}
\me{ \left(\theta^2 -\me{\theta^2} \right)^2}\geq 0\,,  
\end{equation}
whence $\me{\theta^4} - \left( \me{\theta^2}\right)^2 \geq 0$, so that, being  $I_{2,0} = \me{\theta^2}$ and $I_{4,0} = \me{\theta^4}$, we have $I_{4,0} \geq I_{2,0} ^2$. We can thus introduce the parameters $b_1$ and $b_2$ such that
\begin{subequations}
\begin{align}
I_{4,0} &= b_1 I_{2,0}^2\,,\label{eq:I40I20}\\ I_{0,4} &= b_2 I_{0,2}^2\,;
\end{align}
\end{subequations} 
\begin{figure}
    \centering
    \includegraphics[width=0.6\linewidth]{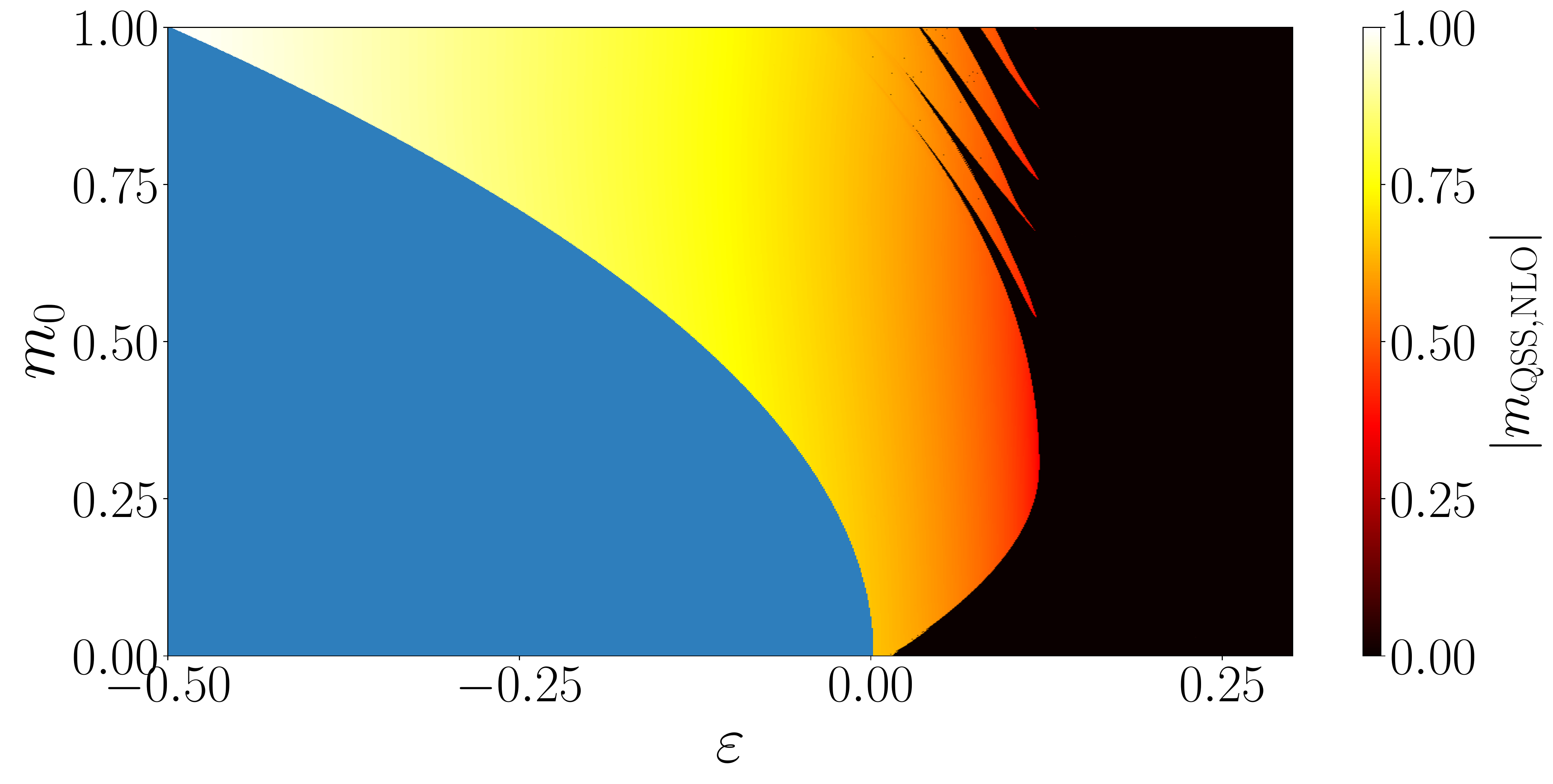}
    \caption{Next-to-leading order prediction for the phase diagram. The parameters are $\gamma_{2} = 0.11$, $b_1=1$ and $b_2=3$ $\gamma_{4}=0.036$. We have taken all the initial odd moments $I_{11},I_{13}$ and $I_{31}$ equal to zero assuming that the initial distribution function can be written as the product of its marginals in $\theta$ and $p$. As in \ref{fig:Theoretical_pd_lo} there are three region in the phase diagram. The warm region is shrinked because the transition to the hot region happens for smaller $\varepsilon$.}
    \label{fig:nlo_theo}
\end{figure}
since we are considering symmetric distributions, $b_1$ and $b_2$ are the kurtosis of the positions and of the velocities, respectively. A uniform distribution has a kurtosis which is equal to $b=1.8$ meanwhile a delta-like distribution has a kurtosis which is equal to $b=1$. In a sense, fixing a value of $b$ means to choose a particular shape of the initial distribution function. However, not every value of $b_1$ allows us to obtain all the possible initial magnetizations $m_0 \in [0,1]$, because Eq.\  \eqref{eq:I40I20} implies
\begin{equation}
m_0 = 1 - \frac{1}{2} I_{2,0}(0) +\frac{1}{24}b_1I_{2,0}^2(0)\,.\label{eq:relation_m0_nlo}
\end{equation}
\begin{figure}
    \centering
    \includegraphics[width=0.5\linewidth]{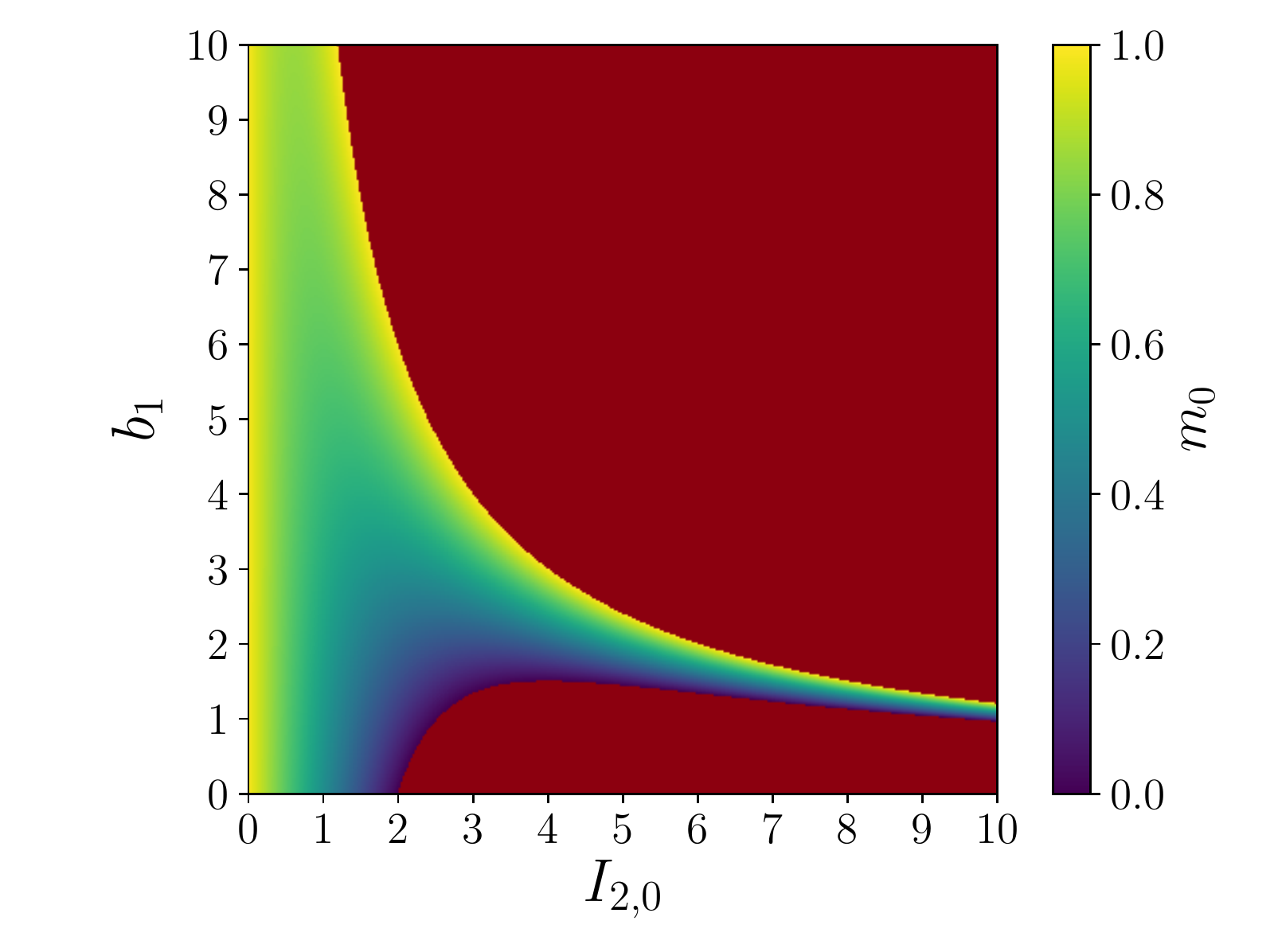}
    \caption{Heatmap of $m_0$ against $I_{2,0}$ and $b_1$ as defined in Eq.\ \eqref{eq:relation_m0_nlo}. The red region is the unphysical one in which $m_0$ would be greater than $1$ or smaller than $0$; it is apparent that for large values of $b_1$ it is not possible to achieve small values of $m_0$.}
    \label{fig:diseq}
\end{figure}
In figure \ref{fig:diseq} we plot $m_0$ as a function $I_{2,0}$ and $b_1$ according to Eq.\ \eqref{eq:relation_m0_nlo}, highlighting the allowed region, i.e., the domain of the $I_{2,0} - b_1$ plane such that $0\leq m_0\leq 1$. In order to explore the whole range of values of $m_0 \in [0,1]$ we have to choose small values of $b_1$. Indeed, given an initial magnetisation $m_0$ we can solve Eq.\ \eqref{eq:relation_m0_nlo} for $I_{2,0}$, obtaining
\begin{equation}
I_{2,0} = \frac{6}{b_1}\left[ 1 \pm \sqrt{1-\frac{2}{3} b_1 (1-m_0)}\right]
\end{equation}
which is a real number only if $b_1 (1-m_0)\leq 3/2$ so that in order to explore all the phase diagram we must keep $b_1 \leq 3/2$, otherwise we would obtain only the region with $m_0>1-3/(2b_1)$. 

In figure \ref{fig:nlo_theo} we plot the non-equilibrium phase diagram at the next-to-leading order, i.e., obtained by solving Eqs. \eqref{eqs:evo_nlo}.

\section{Comparison of the theoretical predictions with numerical results}
\label{sec:comparison}
In order to check our predictions we solved the Vlasov equation for the HMF model using a semi-Lagrangian method \cite{DEBUYL20141822, 1976JChengKnorr, SONNENDRUCKER1999201} using different prescriptions for the initial conditions of the distribution function. 

For the sake of simplicity, and to reduce the huge space of possible initial conditions, we consider an initial distribution function factorized in its marginals, i.e.,
\begin{equation}
    f(\theta,p,t=0) \equiv f_0(\theta,p) = h(\theta)g(p)\,;
\end{equation}
this means that we are restricting ourselves to the particular case in which the initial covariance $\me{\theta p}$ vanishes. Then we tune $h(\theta)$ in order to set the initial magnetization $m_0=\me{\cos\theta}$. Finally, we set the correct variance of $g(p)$, which is twice the initial kinetic energy, in order to obtain the desired value of the initial energy density according to
\begin{equation}
    \me{p^2} = 2 \varepsilon + m_0^2.
\end{equation}
We evolve $f_0(\theta,p)$ up to a time $t_\mathrm{max}$ large enough to be sure that our observable $m$ is stationary. We have found that for all cases considered $t_\mathrm{max}=300$ is a good value. Then we define the quasi-stationary magnetisation $m_\mathrm{QSS}$ as follows: 
\begin{equation}
    m_\mathrm{QSS} = \frac{1}{t_\mathrm{max}-t_0}\int_{t_0}^{t_\mathrm{max}} f(\theta,p,t') \cos\theta \, dt'\,,
\end{equation}
where $t_0=250$, in order to average out the small residual quasi-stationary oscillations. The phase diagrams that we are going to show are plotted on grids whose spacing\footnote{We made sure that the phase-space grids over which the the initial states are defined are dense enough so that the corresponding error on $m_0$ and $\varepsilon$ is negligible compared to $\Delta m_0$ and $\Delta \epsilon$.} is $\Delta \varepsilon = \Delta m_0 = 5 \times 10^{-3}$  in energy density and initial magnetization, respectively. Finally, we compare the numerically obtained values of the quasi-stationary magnetization $m_\mathrm{QSS}$ with our theoretical predictions, and in particular with those reported in Figs.\  \ref{fig:Theoretical_pd_lo}$a)$ and \ref{fig:nlo_theo}. As we shall see in the following, we find an excellent agreement between theory and numerics for all the values of the parameter space but a region close to the transition line. This agreement is somewhat surprising, since our theoretical predictions are a priori expected to be reliable only for initial conditions not too far from cold ones, as discussed above.

The classes of initial conditions we choose to analyze are: (\textit{i}) a waterbag in both momenta and positions (\textit{ii}) a Gaussian in positions and a waterbag in momenta and (\textit{iii}) a Gaussian in both positions and momenta. We shall also report on results obtained from another class of initial conditions which clearly show fringes at the border between phases.

\subsection{Waterbag initial conditions}

\begin{figure}[p]
    \centering
    \includegraphics[width=0.6\linewidth]{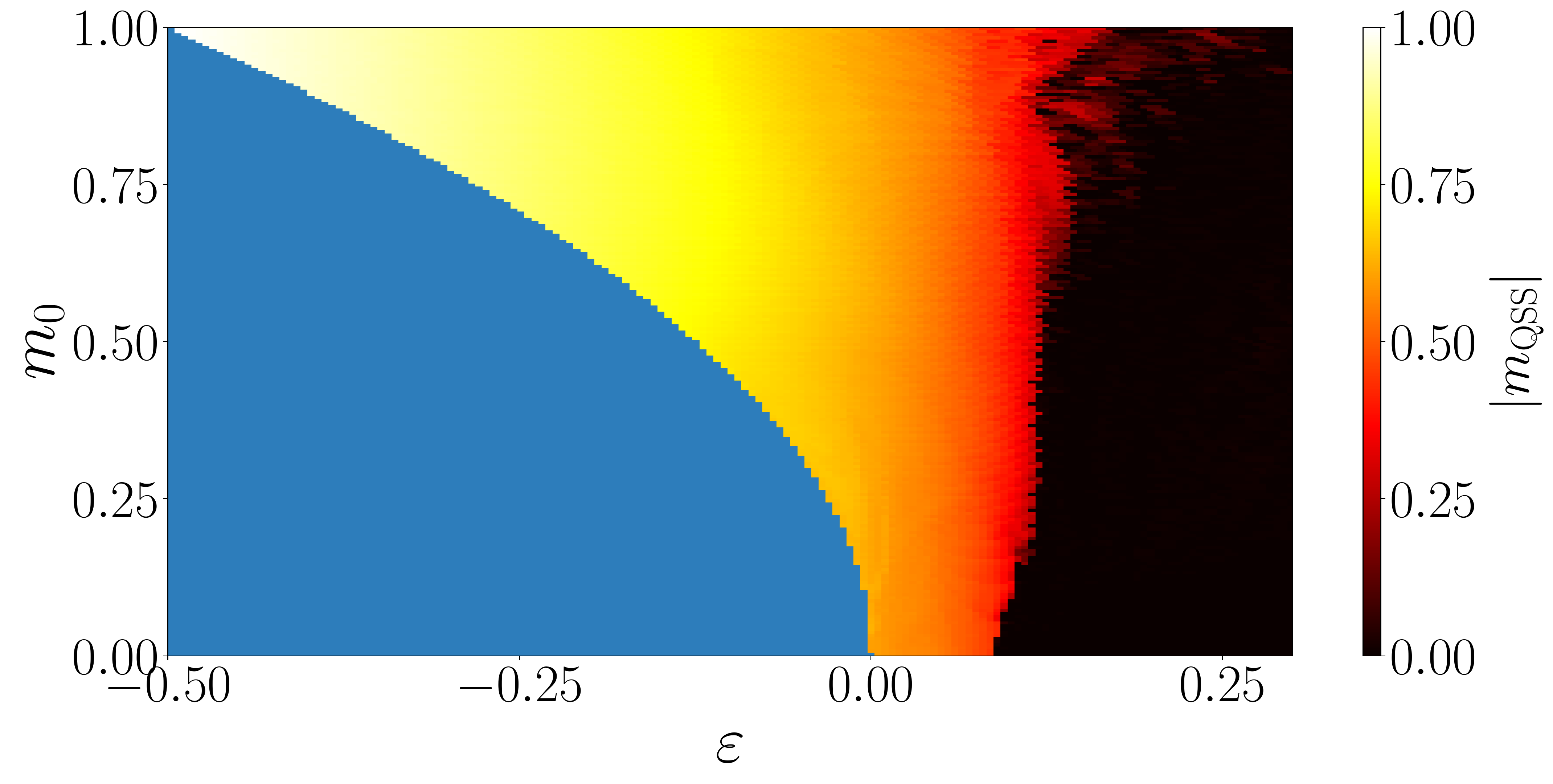}
    \caption{Quasi-stationary magnetization obtained evolving waterbag initial distribution functions with Vlasov simulations.}
    \label{fig:WB_vlasov}
\end{figure}
\begin{figure}[p]
    \centering
    \includegraphics[width=\linewidth]{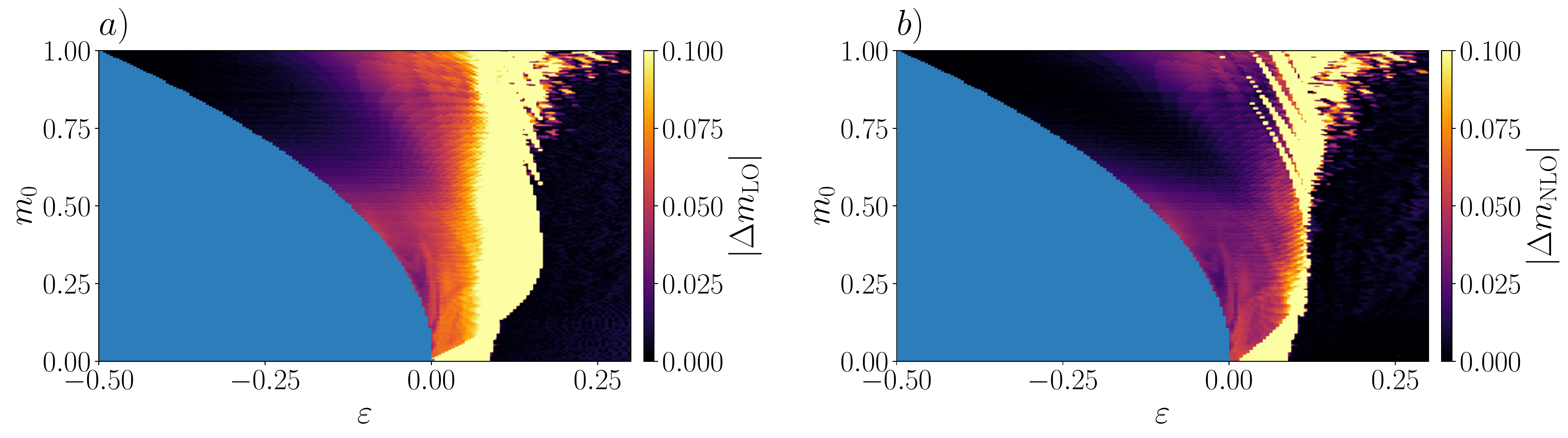}
    \caption{Waterbag initial conditions. Absolute differences $\Delta m_\mathrm{(N)LO} = m_\mathrm{QSS} - m_\mathrm{(N)LO}$ between the Vlasov results shown in Fig.\ \ref{fig:WB_vlasov} and the theoretical predictions. In particular, panels $a)$ and $b)$ show the difference between numerical values and the LO and NLO predictions, respectively.}
    \label{fig:WB_comparison}
\end{figure}
\begin{figure}[p]
    \centering
    \includegraphics[width=.8\linewidth]{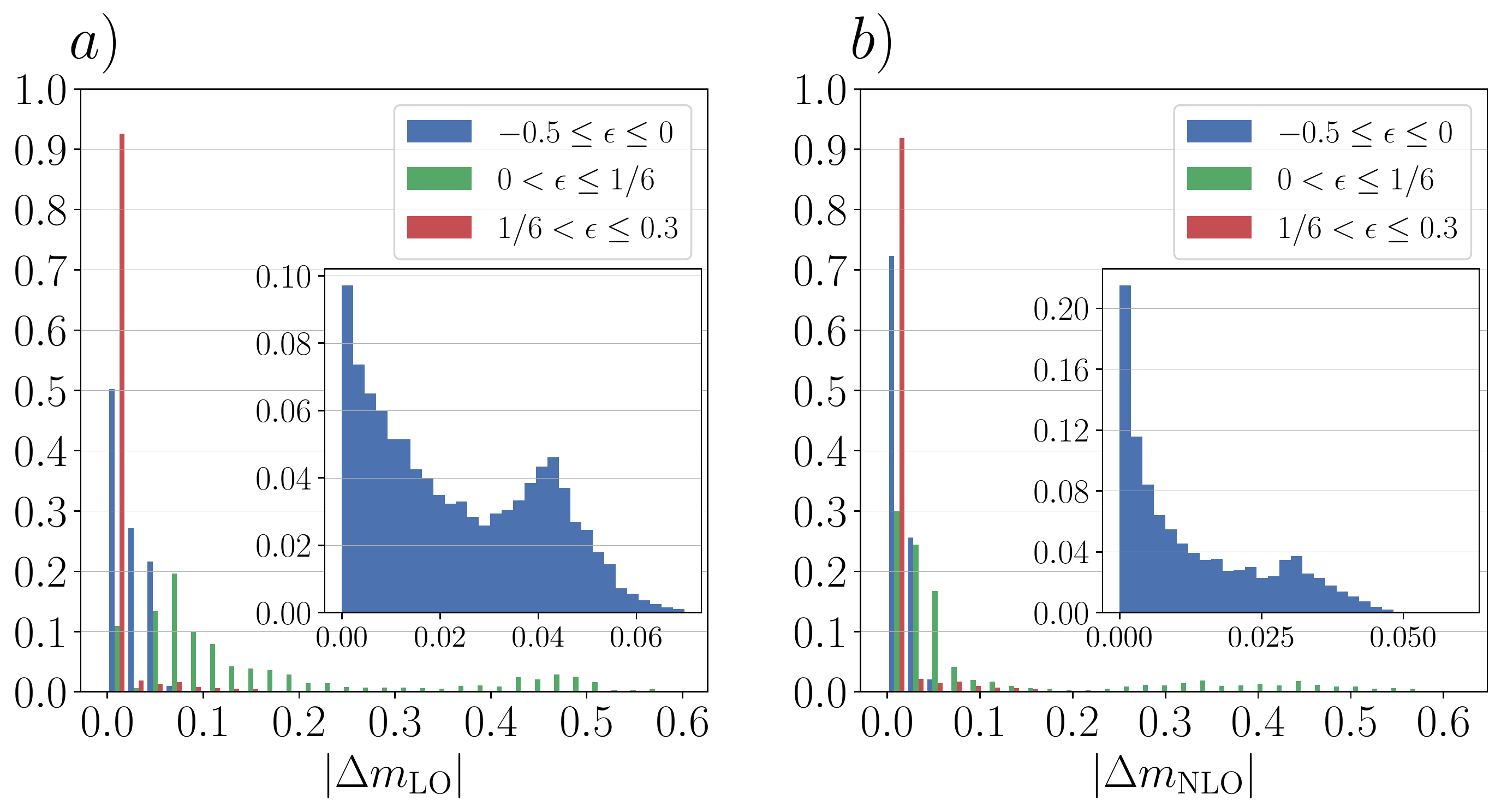}
    \caption{Waterbag initial conditions. Histograms of the absolute differences $\Delta m_\mathrm{(N)LO}$ between the Vlasov results and the theoretical predictions, i.e., of the results shown in Fig.\ \ref{fig:WB_comparison}. We plot the histograms for the cold region $-0.5\leq \varepsilon\leq 0$, the warm region $0<\varepsilon\leq 1/6$ and hot region $\varepsilon>1/6$. In the insets we plot more detailed histograms relative to the cold region.}
    \label{fig:WB_comparison_hist}
\end{figure}

Let us consider initial distribution functions defined as 
\begin{equation}
    f_0(\theta,p)= \frac{1}{4\Delta \theta \Delta p} \Theta(|\theta|-\Delta \theta)\Theta(|p|-\Delta p)\label{eq:InitialWaterbag}
\end{equation}
where $\Theta(\cdot)$ is the Heaviside step function and $\Delta \theta$ and $\Delta p$ are parameters fixed by the desired initial $m_0$ and $\varepsilon$. This kind of ``rectangular'' initial conditions are widely studied in the literature (see e.g.\  \cite{Campa-Dauxois-Fanelli-Ruffo,CampaEtAl:physrep,LevinEtAlphysrep:2014,Antoniazzi2007}) and are commonly referred to as ``waterbag'' initial conditions.

In Figure \ref{fig:WB_vlasov} we plot the resulting non-equilibrium phase diagram, i.e., the values of $|m_\mathrm{QSS}|$ as a function of $m_0$ and $\varepsilon$, for this kind of initial conditions. In Figures \ref{fig:WB_comparison} and \ref{fig:WB_comparison_hist} we compare the latter numerical results with our theoretical prediction at the leading and next-to-leading orders. In particular, in Figure \ref{fig:WB_comparison} we plot the absolute difference between $|m_\mathrm{QSS}|$ and $m_\mathrm{LO}$, panel $a)$, and $m_\mathrm{NLO}$, panel $b)$, against the initial magnetisation $m_0$ and energy density $\varepsilon$, while in Figure \ref{fig:WB_comparison_hist} we plot the histogram of the corresponding absolute differences in the cold, warm and hot regions at LO, panel $a)$, and NLO, panel $b)$. 

We obtain an excellent agreement already at the leading order, apart from a sensible error close to the transition line. Note that in the cold region $\varepsilon<0$ the differences between theory and numerics are really small, as shown in the inset of figure \ref{fig:WB_comparison_hist}$a)$. The next-to-leading order predictions are in even better agreement with the numerics: the error in the $\varepsilon<0$ region is further decreased (as shown in the inset of figure \ref{fig:WB_comparison_hist}$b)$) and the predicted transition line is closer to the correct one. Moreover, there is always a rather sharp boundary between colored and black regions in Fig.\ \ref{fig:WB_vlasov}, thus indicating that the phase transition is discontinuous as predicted by our theoretical approach (although at large values of $m_0$ the phase boundary gets very complicated and it is difficult to draw reliable conclusions on the order of the transition; see also the discussion on Sec.\ \ref{sec:fringes}).  

This particular kind of initial conditions allows us to compare our prediction with other ones known in the literature, and in particular with those based on the seminal work by Lynden-Bell \cite{Lynden-Bell:mnras1967}. Lynden-Bell's theory provides a way to determine the quasi-stationary-state of a long-range interacting system by solving a self-consistent problem. The latter, unfortunately, is (at least partially) analytically treatable only for a very limited choice of initial conditions, including the waterbag distributions. Within this case, Lynden-Bell's theory is known to work pretty well in predicting the phase diagram of the HMF model \cite{Antoniazzi2007}, so that we expect our theory to give comparable results only in the cold region. Our numerical findings confirm this expectation (data not shown): the LO and NLO predictions are very close to Lynden-Bell's in the cold region while are slightly worse in the other regions and especially close to the transition line.

\subsection{Waterbag-Gaussian initial conditions}
\begin{figure}[p]
    \centering
    \includegraphics[width=0.6\linewidth]{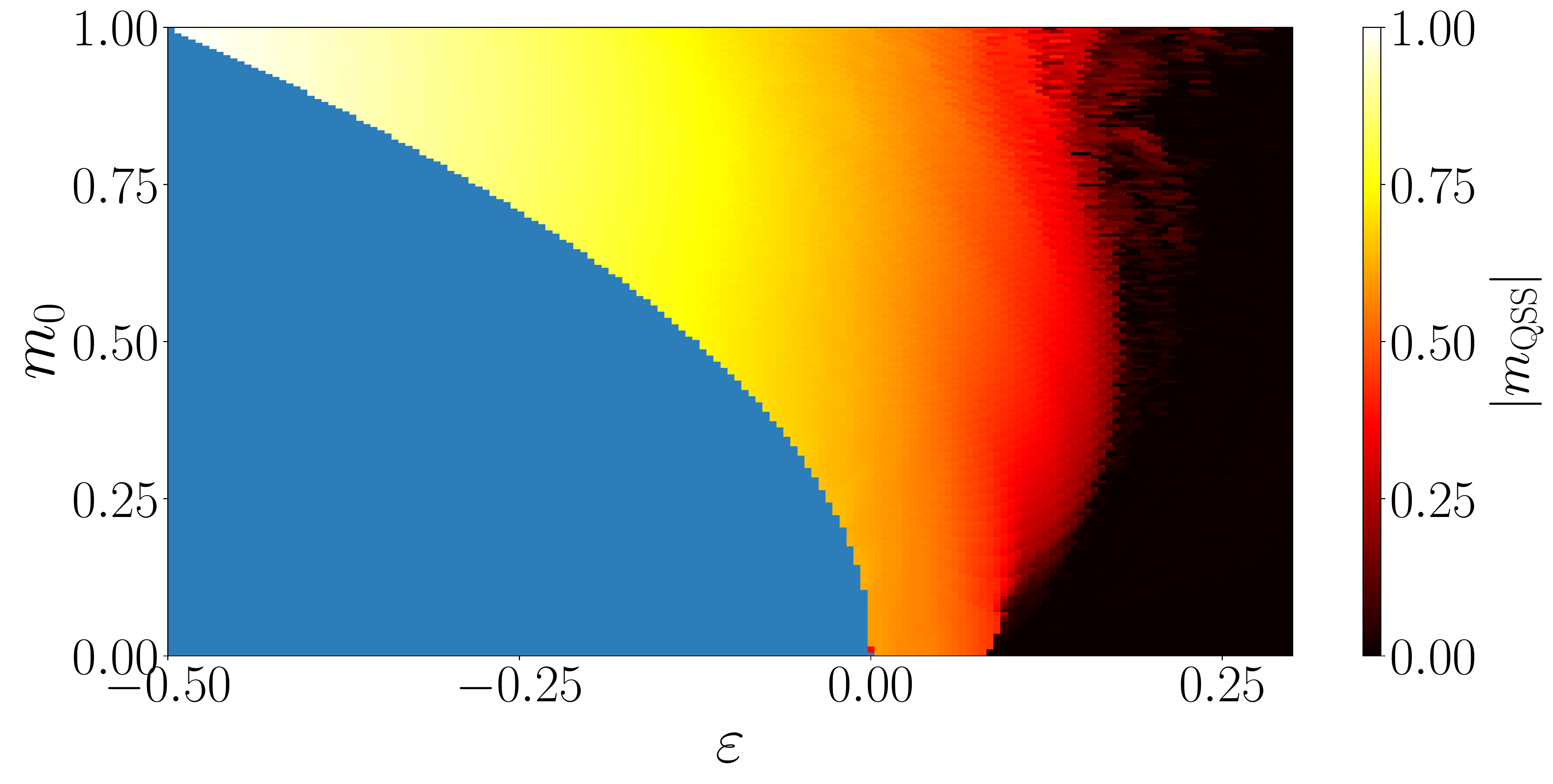}
    \caption{As in Fig.\ \ref{fig:WB_vlasov}, for waterbag-Gaussian initial conditions.}
    \label{fig:GWB_vlasov}
\end{figure}

\begin{figure}[p]
    \centering
    \includegraphics[width=\linewidth]{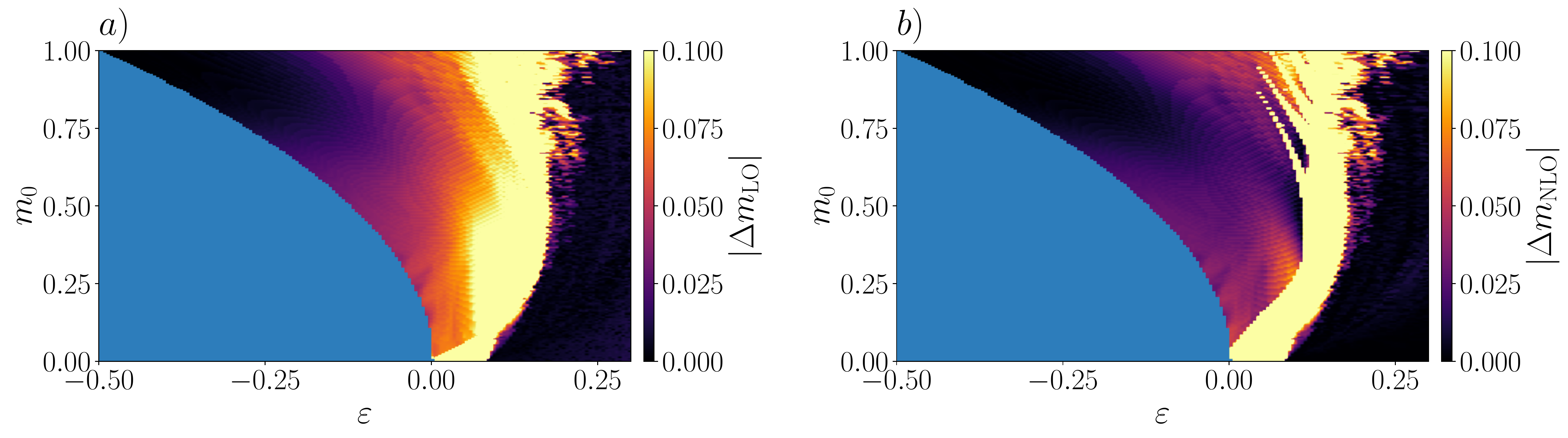}
    \caption{As in Fig.\ \ref{fig:WB_comparison}, for waterbag-Gaussian initial conditions.}
    \label{fig:GWB_comparison}
\end{figure}

\begin{figure}[p]
    \centering
    \includegraphics[width=.8\linewidth]{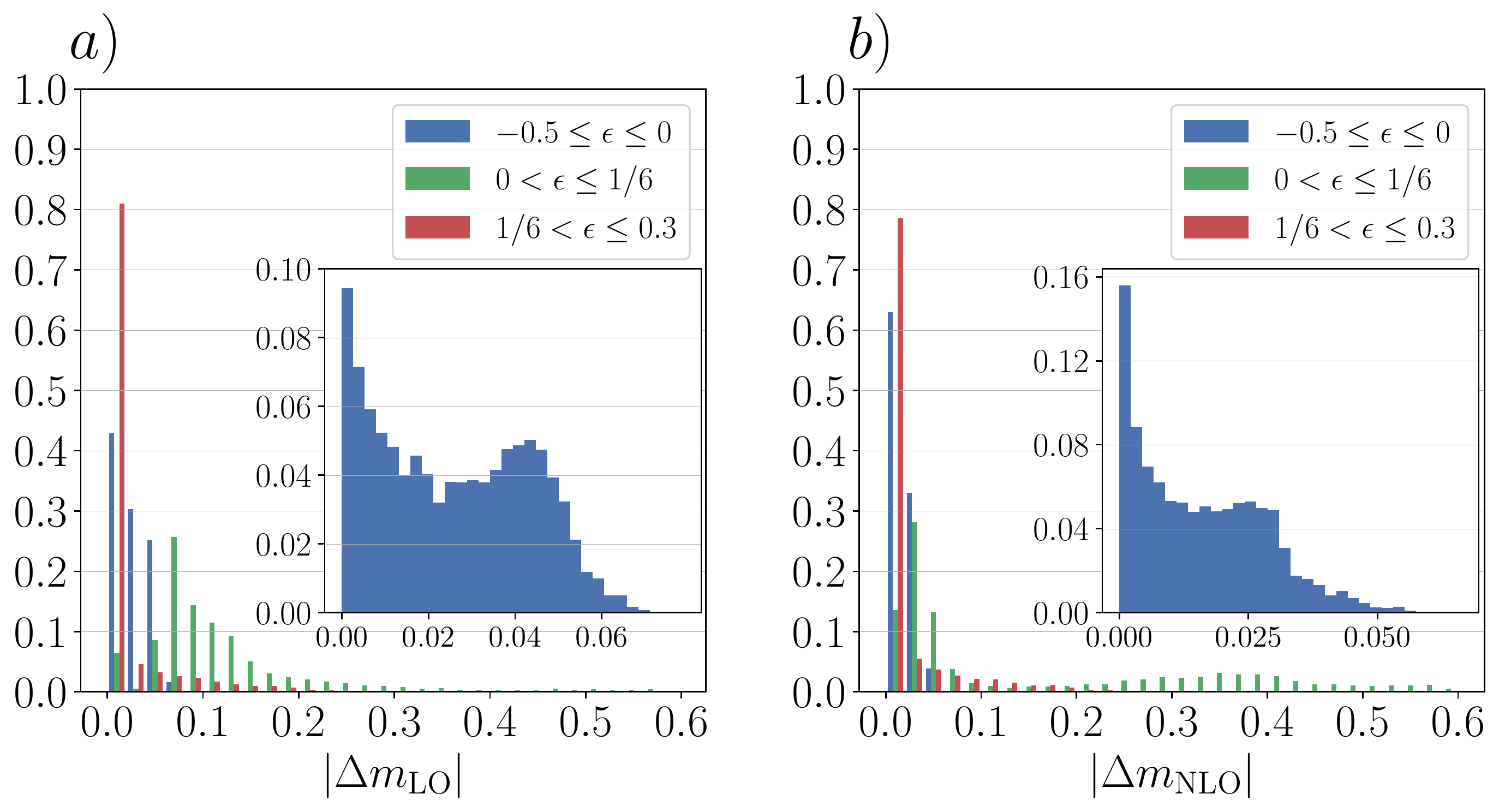}
    \caption{As in Fig.\ \ref{fig:WB_comparison_hist}, for waterbag-Gaussian initial conditions.}
    \label{fig:GWB_comparison_hist}
\end{figure}

Let us now consider the following class of initial conditions:  
\begin{equation}
    f_0(\theta,p) = A(\sigma_\theta) \exp\left(-\frac{\theta^2}{2\sigma_\theta^2}\right)  \frac{1}{2\Delta p}\Theta(|p|-\Delta p)\label{eq:gaussian+waterbag}
\end{equation}
where $A(\sigma_\theta)^{-1} = \sqrt{2 \pi} \sigma_\theta \, \rm erf \left( \frac{\pi}{\sqrt{2} \sigma_\theta} \right)$ provides the normalization while the values of $m_0$ and $\varepsilon$ are determined by varying $\sigma_\theta$ and $\Delta p$, respectively. In Figure \ref{fig:GWB_vlasov} we plot the resulting non-equilibrium phase diagram for this kind of initial conditions and in Figs.\  \ref{fig:GWB_comparison} and \ref{fig:GWB_comparison_hist} we compare our theoretical predictions at the leading and next-to-leading orders to the numerical results. The results are very similar to the the previous case: we obtain a very good agreement between theory and numerics everywhere but for the transition region and the NLO approximation greatly improves the prediction w.r.t. to the LO one within the magnetized region. However, the theory still fails to predict the correct shape and location of the phase boundary. As for waterbag initial conditions, Moreover, the phase transition looks discontinuous as predicted by our theoretical approach at least for sufficiently small values of $m_0$.  



\subsection{Gaussian-Gaussian initial conditions}
\begin{figure}[p]
    \centering
    \includegraphics[width=0.6\linewidth]{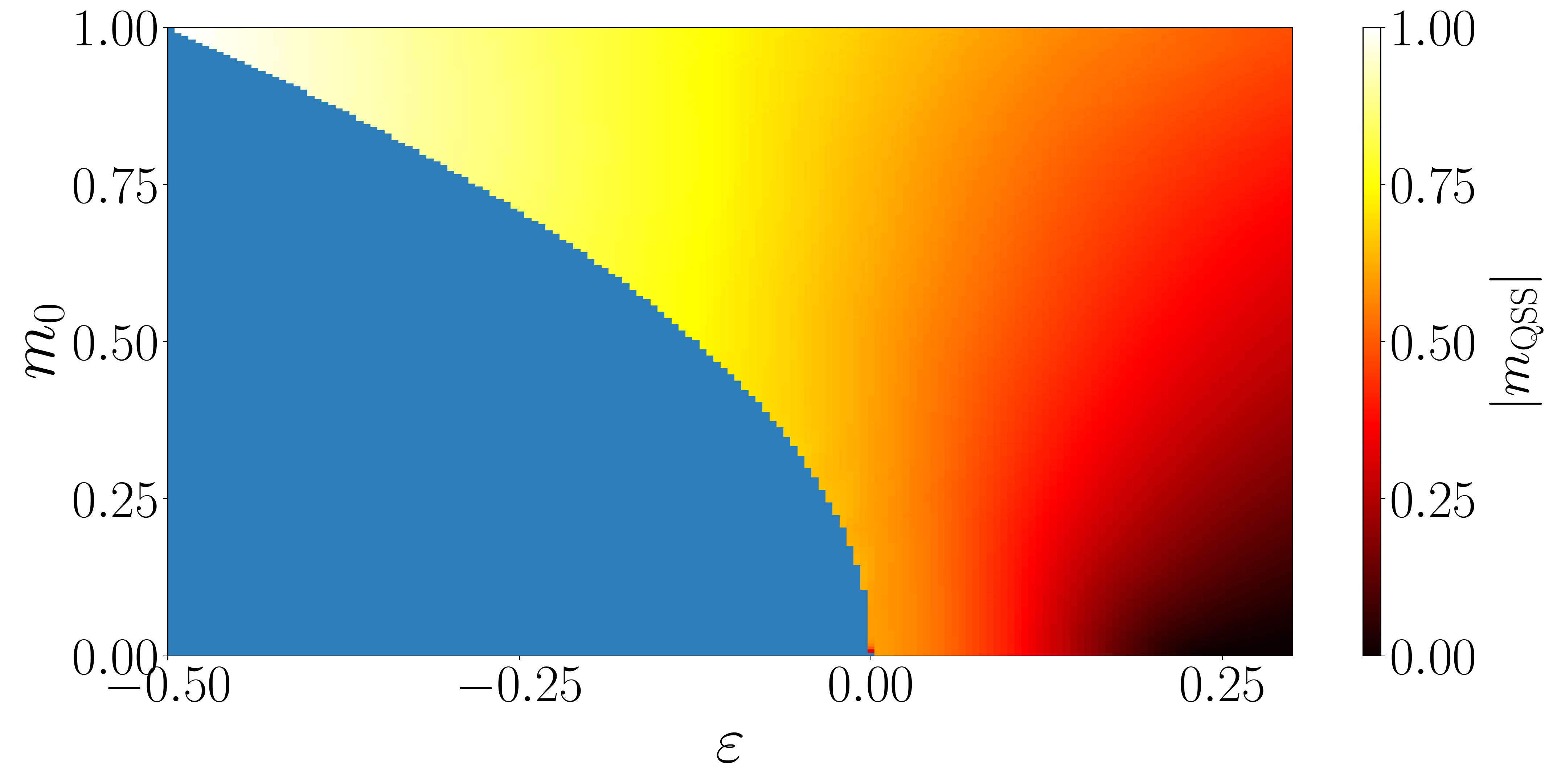}
    \caption{As in Fig.\ \ref{fig:WB_vlasov}, for Gaussian-Gaussian initial conditions.}
    \label{fig:GG_vlasov}
\end{figure}

\begin{figure}[p]
    \centering
    \includegraphics[width=\linewidth]{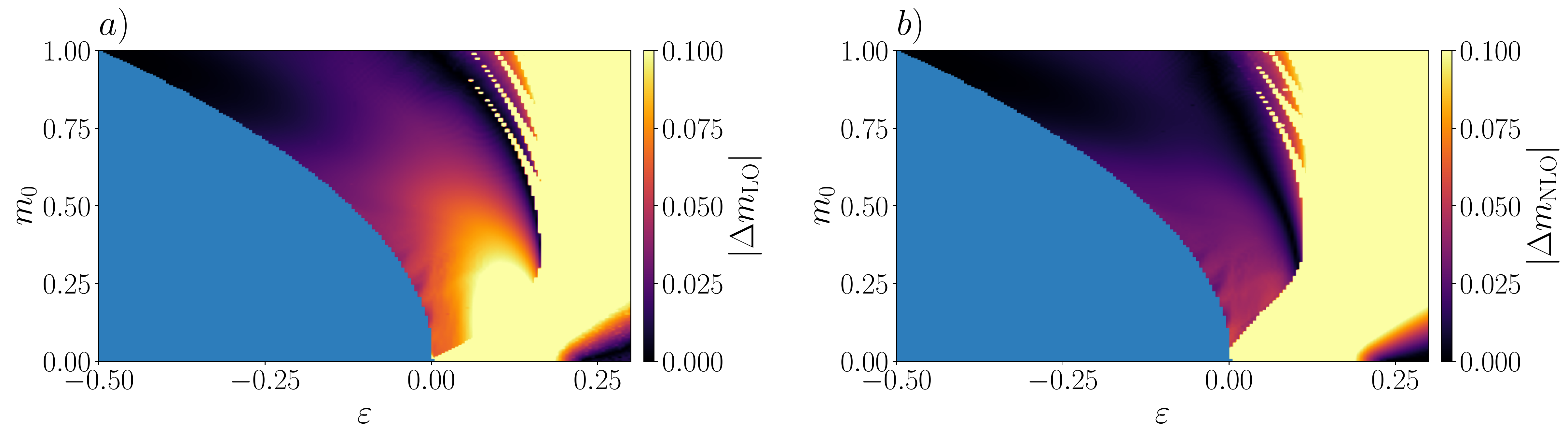}
    \caption{As in Fig.\ \ref{fig:WB_comparison}, for Gaussian-Gaussian initial conditions.}
    \label{fig:GG_comparison}
\end{figure}

\begin{figure}[p]
    \centering
    \includegraphics[width=.8\linewidth]{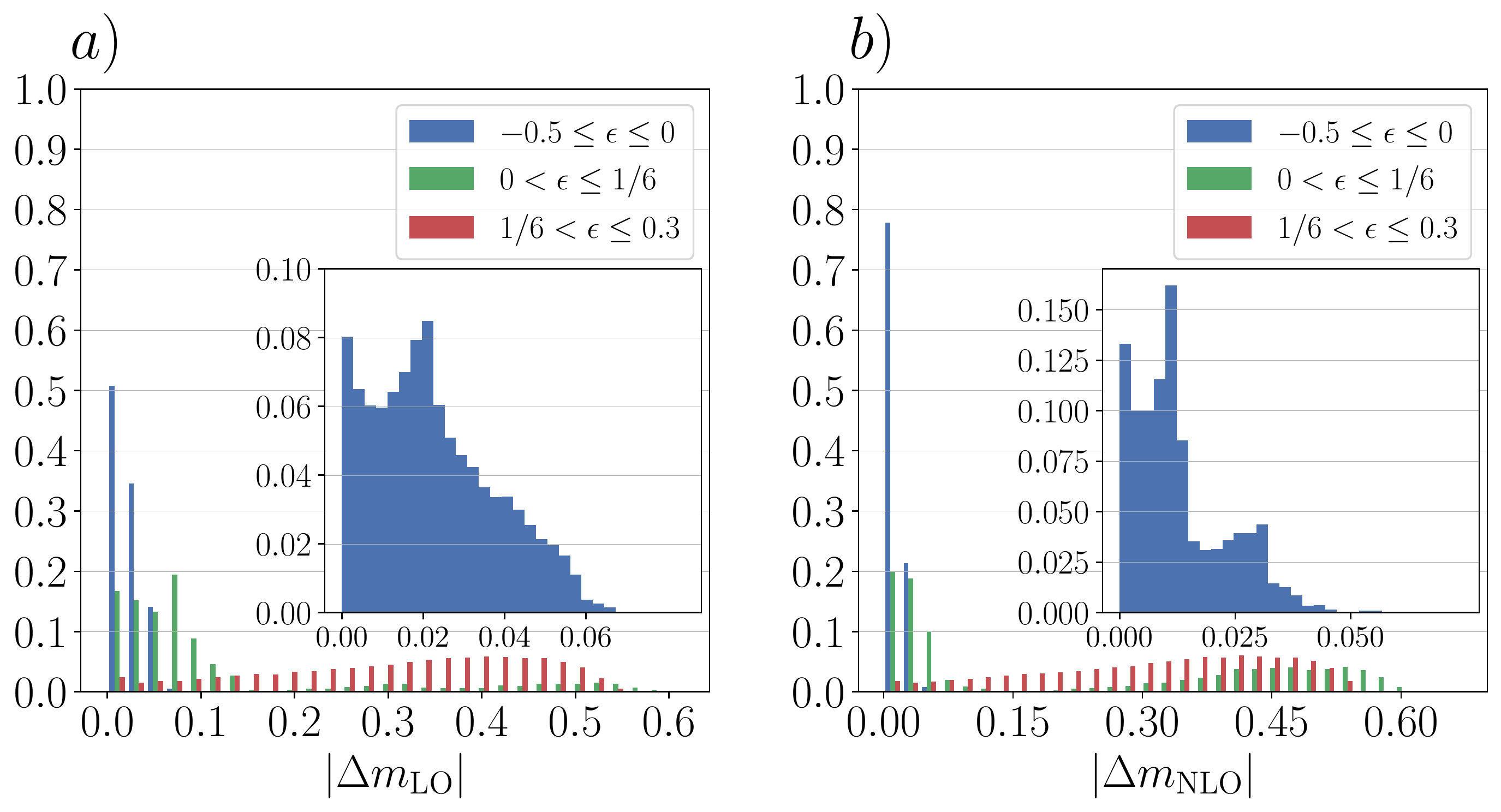}
    \caption{As in Fig.\ \ref{fig:WB_comparison_hist}, for Gaussian-Gaussian initial conditions.}
    \label{fig:GG_comparison_hist}
\end{figure}

We now turn to considering a class of initial conditions such that the distribution function $f_0$ is a Gaussian in both position and momenta, that is,
\begin{equation}
    f_0(\theta,p) = \frac{A(\sigma_\theta)}{\sqrt{2 \pi \sigma_p}} \exp\left[-\left(\frac{\theta^2}{2\sigma_\theta^2}+\frac{p^2}{2\sigma_p^2}\right)\right]~, \label{eq:gaussian+gaussian}
\end{equation}
where, once again, $A(\sigma_\theta)^{-1} = \sqrt{2 \pi} \sigma_\theta \, \rm erf \left( \frac{\pi}{\sqrt{2} \sigma_\theta} \right)$ and the values of $m_0$ and $\varepsilon$ are determined by varying $\sigma_\theta$ and $\Delta p$, respectively
Analogously to the previous cases, in Figure \ref{fig:GG_vlasov} we plot the resulting non-equilibrium phase diagram for this kind of initial conditions and in Figs.\ \ref{fig:GG_comparison} and \ref{fig:GG_comparison_hist} we compare the numerical results with our theoretical prediction at the leading and next-to-leading orders.
As before, the results within the magnetized phase at LO and NLO are really good. In this case, however, according to the Vlasov simulation the phase transition to the paramagnetic phase happens at higher values of $\varepsilon$ (see Fig.\ \ref{fig:GG_vlasov}). The latter effect might be due to the fact that especially when $m_0$ approaches $1$ the initial shape of the distribution function is closer and closer to a collapsed thermal equilibrium distribution. This may imply that the system is more stable and needs more kinetic energy in order to escape the collapsed state. Moreover, the transition now appears as continuous, as it happens in thermal equilibrium. In any case, these features cannot be predicted by our theory.



\subsection{Numerical evidence of fringes in the phase boundary}
\label{sec:fringes}

\begin{figure}
    \centering
    \includegraphics[width=0.6\linewidth]{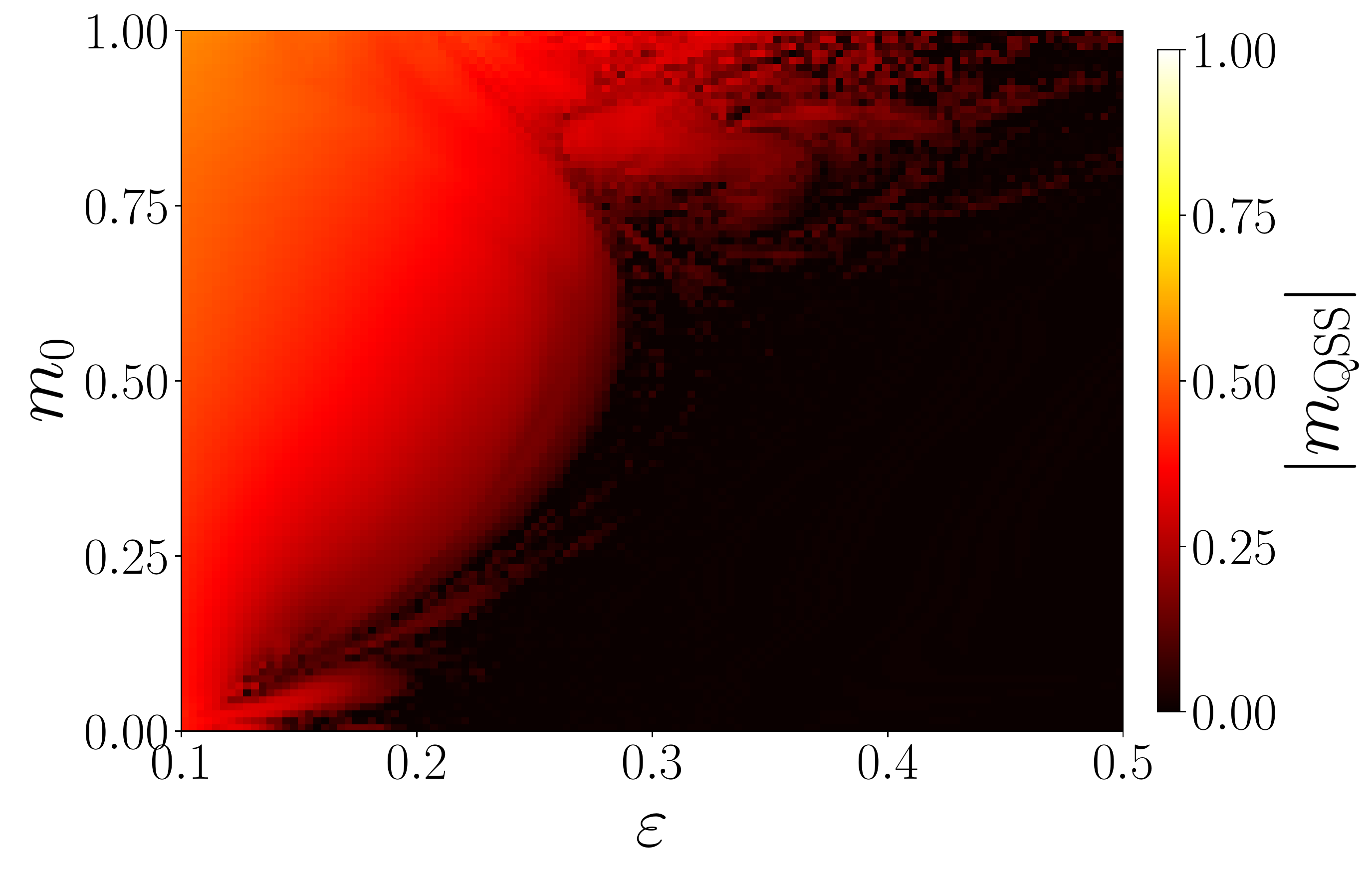}
    \caption{Detail of the phase boundary as obtained solving the Vlasov equation from initial conditions of the class \eqref{eq:distribuzione_strana}. Fringes in the phase boundary, implying reentrant phase transition patterns when moving either vertically or horizontally in the phase diagram, are apparent for $\varepsilon  < 0.3$.}
    \label{fig:detail}
\end{figure}

In spite of the fact that the shape of the transition line close to $m_0=1$ in Figs.\ \ref{fig:WB_vlasov} and \ref{fig:GWB_vlasov} is pretty complicated, so far one may think the neat fringes (implying a reentrant behavior) in the phase diagram predicted by our theory are an artifact. However, as we are going to show, there also are initial conditions such that neat fringes do appear in the phase boundary, which are qualitatively very similar to the theoretically predicted ones. As an example, let us consider a class of initial conditions still factorized, i.e., such that $f_0(\theta,p)=g(\theta)h(p)$, with
\begin{subequations}\begin{align}
g(\theta) & = A(\sigma_\theta) \exp\left[-\frac{1}{2}\left(\frac{\theta}{\sigma_\theta}\right)^2\right], \\
h(p) & =\frac{\lambda^3}{4}p^2e^{-\lambda |p|},
\end{align}\label{eq:distribuzione_strana}
\end{subequations}
where, as before,   $A(\sigma_\theta)^{-1} = \sqrt{2 \pi} \sigma_\theta \, \rm erf \left( \frac{\pi}{\sqrt{2} \sigma_\theta} \right)$ while $\sigma_\theta$ and $\lambda$ fix $m_0$ and $\varepsilon$, respectively. In Fig.\ \ref{fig:detail} we show a detail of the non-equilibrium phase diagram, corresponding to $\varepsilon \in [0.1,0.5]$ and $m_0 \in [0,1]$, obtained with a finer grid, i.e., $\Delta m_0 = 10^{-2}$ and $\Delta \varepsilon = 3 \times 10^{-3}$. The transition line looks complicated and fractal-like; this notwithstanding, a clear concave shape, like the ones we predict with our theory, is clearly visible for $\varepsilon  < 0.3$. 

\section{Concluding remarks}
\label{sec:concl}
We have shown that the approximate treatment of violent relaxation in the HMF model based on a truncated hierarchy of moments of the distribution function, which had been introduced in Paper I for cold initial conditions, can be generalized to generic initial conditions. We have explicitly implemented the approximation at the leading and at the next-to-leading order, producing theoretical non-equilibrium phase diagrams (or more precisely ``3d'' phase diagrams where not only the phase boundary but also the actual value of the order parameter is shown). The latter have been compared to Vlasov numerical simulations carried out starting from four different classes of initial conditions. The agreement between theoretical predictions and numerical results is typically very good over the entire phase diagram but for a region around the phase transition, and is way beyond naive expectations, given that the theory was put forward to describe only collapsed phases and cold initial conditions. Our theoretical approach fails in quantitatively describing the position and the shape of the phase boundary between ferromagnetic and paramagnetic regions, but correctly predicts the qualitative features of the transition. In most of the ferromagnetic phase the difference between theoretical and numerical values of the quasi-stationary magnetization is very small. Moreover, the agreement between theory and numerics always improves, often considerably, going from the leading to the next-to-leading order, and the width of the region around the phase boundary where the theory quantitatively fails shrinks. This suggests that going to higher order should allow to obtain even better results. Higher orders would also allow to more efficiently predict how the quasi-stationary values of the magnetization depend on the kind of initial condition, and not only on the initial values of the energy and of the magnetization. Indeed, at the LO the theoretical prediction only depends on $\varepsilon$ and $m_0$, while at the NLO a further dependence on a single parameter related to the shape of the initial distribution function appears. At higher orders one expects to be able to better resolve different classes of initial conditions, because results would depend on a larger number of parameters in addition to $\varepsilon$ and $m_0$. However, also the number of relevant moments (and thus the number of differential equations to be derived and solved) grows from three to eight going from LO to NLO and would rapidly increase at higher orders, as well as the number of free parameters, i.e., the damping coefficients of the of $p$-odd moments.  

\acknowledgments

This work is part of MIUR-PRIN2017 Project No.\ 201798CZL,``Coarse-grained description for non-equilibrium systems and transport phenomena (CO-NEST)'', whose partial financial support is acknowledged.

\bibliography{bibliography}

\begin{thebibliography}{22}%
\makeatletter
\providecommand \@ifxundefined [1]{%
 \@ifx{#1\undefined}
}%
\providecommand \@ifnum [1]{%
 \ifnum #1\expandafter \@firstoftwo
 \else \expandafter \@secondoftwo
 \fi
}%
\providecommand \@ifx [1]{%
 \ifx #1\expandafter \@firstoftwo
 \else \expandafter \@secondoftwo
 \fi
}%
\providecommand \natexlab [1]{#1}%
\providecommand \enquote  [1]{``#1''}%
\providecommand \bibnamefont  [1]{#1}%
\providecommand \bibfnamefont [1]{#1}%
\providecommand \citenamefont [1]{#1}%
\providecommand \href@noop [0]{\@secondoftwo}%
\providecommand \href [0]{\begingroup \@sanitize@url \@href}%
\providecommand \@href[1]{\@@startlink{#1}\@@href}%
\providecommand \@@href[1]{\endgroup#1\@@endlink}%
\providecommand \@sanitize@url [0]{\catcode `\\12\catcode `\$12\catcode
  `\&12\catcode `\#12\catcode `\^12\catcode `\_12\catcode `\%12\relax}%
\providecommand \@@startlink[1]{}%
\providecommand \@@endlink[0]{}%
\providecommand \url  [0]{\begingroup\@sanitize@url \@url }%
\providecommand \@url [1]{\endgroup\@href {#1}{\urlprefix }}%
\providecommand \urlprefix  [0]{URL }%
\providecommand \Eprint [0]{\href }%
\providecommand \doibase [0]{http://dx.doi.org/}%
\providecommand \selectlanguage [0]{\@gobble}%
\providecommand \bibinfo  [0]{\@secondoftwo}%
\providecommand \bibfield  [0]{\@secondoftwo}%
\providecommand \translation [1]{[#1]}%
\providecommand \BibitemOpen [0]{}%
\providecommand \bibitemStop [0]{}%
\providecommand \bibitemNoStop [0]{.\EOS\space}%
\providecommand \EOS [0]{\spacefactor3000\relax}%
\providecommand \BibitemShut  [1]{\csname bibitem#1\endcsname}%
\let\auto@bib@innerbib\@empty
\bibitem [{\citenamefont {Giachetti}\ and\ \citenamefont
  {Casetti}(2019)}]{Giachetti_2019}%
  \BibitemOpen
  \bibfield  {author} {\bibinfo {author} {\bibfnamefont {G.}~\bibnamefont
  {Giachetti}}\ and\ \bibinfo {author} {\bibfnamefont {L.}~\bibnamefont
  {Casetti}},\ }\href {\doibase 10.1088/1742-5468/ab0c19} {\bibfield  {journal}
  {\bibinfo  {journal} {Journal of Statistical Mechanics: Theory and
  Experiment}\ }\textbf {\bibinfo {volume} {2019}},\ \bibinfo {pages} {043201}
  (\bibinfo {year} {2019})}\BibitemShut {NoStop}%
\bibitem [{\citenamefont {Latella}\ \emph {et~al.}(2015)\citenamefont
  {Latella}, \citenamefont {P\'erez-Madrid}, \citenamefont {Campa},
  \citenamefont {Casetti},\ and\ \citenamefont {Ruffo}}]{prl2015}%
  \BibitemOpen
  \bibfield  {author} {\bibinfo {author} {\bibfnamefont {I.}~\bibnamefont
  {Latella}}, \bibinfo {author} {\bibfnamefont {A.}~\bibnamefont
  {P\'erez-Madrid}}, \bibinfo {author} {\bibfnamefont {A.}~\bibnamefont
  {Campa}}, \bibinfo {author} {\bibfnamefont {L.}~\bibnamefont {Casetti}}, \
  and\ \bibinfo {author} {\bibfnamefont {S.}~\bibnamefont {Ruffo}},\ }\href
  {\doibase 10.1103/PhysRevLett.114.230601} {\bibfield  {journal} {\bibinfo
  {journal} {Phys. Rev. Lett.}\ }\textbf {\bibinfo {volume} {114}},\ \bibinfo
  {pages} {230601} (\bibinfo {year} {2015})}\BibitemShut {NoStop}%
\bibitem [{\citenamefont {{Gupta}}\ and\ \citenamefont
  {{Casetti}}(2016)}]{njp2016}%
  \BibitemOpen
  \bibfield  {author} {\bibinfo {author} {\bibfnamefont {S.}~\bibnamefont
  {{Gupta}}}\ and\ \bibinfo {author} {\bibfnamefont {L.}~\bibnamefont
  {{Casetti}}},\ }\href {\doibase 10.1088/1367-2630/18/10/103051} {\bibfield
  {journal} {\bibinfo  {journal} {New Journal of Physics}\ }\textbf {\bibinfo
  {volume} {18}},\ \bibinfo {eid} {103051} (\bibinfo {year}
  {2016})}\BibitemShut {NoStop}%
\bibitem [{\citenamefont {Campa}\ \emph {et~al.}(2014)\citenamefont {Campa},
  \citenamefont {Dauxois}, \citenamefont {Fanelli},\ and\ \citenamefont
  {Ruffo}}]{Campa-Dauxois-Fanelli-Ruffo}%
  \BibitemOpen
  \bibfield  {author} {\bibinfo {author} {\bibfnamefont {A.}~\bibnamefont
  {Campa}}, \bibinfo {author} {\bibfnamefont {T.}~\bibnamefont {Dauxois}},
  \bibinfo {author} {\bibfnamefont {D.}~\bibnamefont {Fanelli}}, \ and\
  \bibinfo {author} {\bibfnamefont {S.}~\bibnamefont {Ruffo}},\ }\href
  {\doibase doi: 10.1093/acprof:oso/9780199581931.001.0001} {\emph {\bibinfo
  {title} {Physics of Long-Range Interacting Systems}}}\ (\bibinfo  {publisher}
  {Oxford University Press},\ \bibinfo {year} {2014})\BibitemShut {NoStop}%
\bibitem [{\citenamefont {Campa}\ \emph {et~al.}(2009)\citenamefont {Campa},
  \citenamefont {Dauxois},\ and\ \citenamefont {Ruffo}}]{CampaEtAl:physrep}%
  \BibitemOpen
  \bibfield  {author} {\bibinfo {author} {\bibfnamefont {A.}~\bibnamefont
  {Campa}}, \bibinfo {author} {\bibfnamefont {T.}~\bibnamefont {Dauxois}}, \
  and\ \bibinfo {author} {\bibfnamefont {S.}~\bibnamefont {Ruffo}},\ }\href
  {\doibase 10.1016/j.physrep.2009.07.001} {\bibfield  {journal} {\bibinfo
  {journal} {Physics Reports}\ }\textbf {\bibinfo {volume} {480}},\ \bibinfo
  {pages} {57} (\bibinfo {year} {2009})}\BibitemShut {NoStop}%
\bibitem [{\citenamefont {Levin}\ \emph {et~al.}(2014)\citenamefont {Levin},
  \citenamefont {Pakter}, \citenamefont {Rizzato}, \citenamefont {Teles},\ and\
  \citenamefont {Benetti}}]{LevinEtAlphysrep:2014}%
  \BibitemOpen
  \bibfield  {author} {\bibinfo {author} {\bibfnamefont {Y.}~\bibnamefont
  {Levin}}, \bibinfo {author} {\bibfnamefont {R.}~\bibnamefont {Pakter}},
  \bibinfo {author} {\bibfnamefont {F.~B.}\ \bibnamefont {Rizzato}}, \bibinfo
  {author} {\bibfnamefont {T.~N.}\ \bibnamefont {Teles}}, \ and\ \bibinfo
  {author} {\bibfnamefont {F.~P.~C.}\ \bibnamefont {Benetti}},\ }\href
  {\doibase http://dx.doi.org/10.1016/j.physrep.2013.10.001} {\bibfield
  {journal} {\bibinfo  {journal} {Physics Reports}\ }\textbf {\bibinfo {volume}
  {535}},\ \bibinfo {pages} {1 } (\bibinfo {year} {2014})}\BibitemShut
  {NoStop}%
\bibitem [{\citenamefont {Giachetti}\ \emph
  {et~al.}(2021{\natexlab{a}})\citenamefont {Giachetti}, \citenamefont
  {Defenu}, \citenamefont {Ruffo},\ and\ \citenamefont
  {Trombettoni}}]{eps2021}%
  \BibitemOpen
  \bibfield  {author} {\bibinfo {author} {\bibfnamefont {G.}~\bibnamefont
  {Giachetti}}, \bibinfo {author} {\bibfnamefont {N.}~\bibnamefont {Defenu}},
  \bibinfo {author} {\bibfnamefont {S.}~\bibnamefont {Ruffo}}, \ and\ \bibinfo
  {author} {\bibfnamefont {A.}~\bibnamefont {Trombettoni}},\ }\href {\doibase
  10.1209/0295-5075/133/57004} {\bibfield  {journal} {\bibinfo  {journal}
  {EPL}\ }\textbf {\bibinfo {volume} {133}},\ \bibinfo {pages} {57004}
  (\bibinfo {year} {2021}{\natexlab{a}})}\BibitemShut {NoStop}%
\bibitem [{\citenamefont {Giachetti}\ \emph
  {et~al.}(2021{\natexlab{b}})\citenamefont {Giachetti}, \citenamefont
  {Defenu}, \citenamefont {Ruffo},\ and\ \citenamefont
  {Trombettoni}}]{Giachetti2021}%
  \BibitemOpen
  \bibfield  {author} {\bibinfo {author} {\bibfnamefont {G.}~\bibnamefont
  {Giachetti}}, \bibinfo {author} {\bibfnamefont {N.}~\bibnamefont {Defenu}},
  \bibinfo {author} {\bibfnamefont {S.}~\bibnamefont {Ruffo}}, \ and\ \bibinfo
  {author} {\bibfnamefont {A.}~\bibnamefont {Trombettoni}},\ }\href@noop {} {\
  (\bibinfo {year} {2021}{\natexlab{b}})},\ \Eprint
  {http://arxiv.org/abs/2104.13217} {arXiv:2104.13217 [cond-mat.stat-mech]}
  \BibitemShut {NoStop}%
\bibitem [{\citenamefont {Teles}\ \emph {et~al.}(2015)\citenamefont {Teles},
  \citenamefont {Gupta}, \citenamefont {Di~Cintio},\ and\ \citenamefont
  {Casetti}}]{prerap2015}%
  \BibitemOpen
  \bibfield  {author} {\bibinfo {author} {\bibfnamefont {T.~N.}\ \bibnamefont
  {Teles}}, \bibinfo {author} {\bibfnamefont {S.}~\bibnamefont {Gupta}},
  \bibinfo {author} {\bibfnamefont {P.}~\bibnamefont {Di~Cintio}}, \ and\
  \bibinfo {author} {\bibfnamefont {L.}~\bibnamefont {Casetti}},\ }\href
  {\doibase 10.1103/PhysRevE.92.020101} {\bibfield  {journal} {\bibinfo
  {journal} {Phys. Rev. E}\ }\textbf {\bibinfo {volume} {92}},\ \bibinfo
  {pages} {020101} (\bibinfo {year} {2015})}\BibitemShut {NoStop}%
\bibitem [{\citenamefont {{Di Cintio}}\ \emph {et~al.}(2018)\citenamefont {{Di
  Cintio}}, \citenamefont {{Gupta}},\ and\ \citenamefont
  {{Casetti}}}]{mnras2018}%
  \BibitemOpen
  \bibfield  {author} {\bibinfo {author} {\bibfnamefont {P.}~\bibnamefont {{Di
  Cintio}}}, \bibinfo {author} {\bibfnamefont {S.}~\bibnamefont {{Gupta}}}, \
  and\ \bibinfo {author} {\bibfnamefont {L.}~\bibnamefont {{Casetti}}},\ }\href
  {\doibase 10.1093/mnras/stx3244} {\bibfield  {journal} {\bibinfo  {journal}
  {Mon. Not. Royal Astron. Soc.}\ }\textbf {\bibinfo {volume} {475}},\ \bibinfo
  {pages} {1137} (\bibinfo {year} {2018})}\BibitemShut {NoStop}%
\bibitem [{\citenamefont {Lynden-Bell}(1967)}]{Lynden-Bell:mnras1967}%
  \BibitemOpen
  \bibfield  {author} {\bibinfo {author} {\bibfnamefont {D.}~\bibnamefont
  {Lynden-Bell}},\ }\href {\doibase 10.1093/mnras/136.1.101} {\bibfield
  {journal} {\bibinfo  {journal} {Mon. Not. Royal Astron. Soc.}\ }\textbf
  {\bibinfo {volume} {136}},\ \bibinfo {pages} {101} (\bibinfo {year}
  {1967})}\BibitemShut {NoStop}%
\bibitem [{\citenamefont {Giachetti}\ \emph {et~al.}(2020)\citenamefont
  {Giachetti}, \citenamefont {Santini},\ and\ \citenamefont
  {Casetti}}]{PhysRevResearch.2.023379}%
  \BibitemOpen
  \bibfield  {author} {\bibinfo {author} {\bibfnamefont {G.}~\bibnamefont
  {Giachetti}}, \bibinfo {author} {\bibfnamefont {A.}~\bibnamefont {Santini}},
  \ and\ \bibinfo {author} {\bibfnamefont {L.}~\bibnamefont {Casetti}},\ }\href
  {\doibase 10.1103/PhysRevResearch.2.023379} {\bibfield  {journal} {\bibinfo
  {journal} {Phys. Rev. Research}\ }\textbf {\bibinfo {volume} {2}},\ \bibinfo
  {pages} {023379} (\bibinfo {year} {2020})}\BibitemShut {NoStop}%
\bibitem [{\citenamefont {Barr\'{e}}\ \emph {et~al.}(2011)\citenamefont
  {Barr\'{e}}, \citenamefont {Olivetti},\ and\ \citenamefont
  {Yamaguchi}}]{BarreOlivettiYamaguchi:jphysa2011}%
  \BibitemOpen
  \bibfield  {author} {\bibinfo {author} {\bibfnamefont {J.}~\bibnamefont
  {Barr\'{e}}}, \bibinfo {author} {\bibfnamefont {A.}~\bibnamefont {Olivetti}},
  \ and\ \bibinfo {author} {\bibfnamefont {Y.~Y.}\ \bibnamefont {Yamaguchi}},\
  }\href {\doibase 10.1088/1751-8113/44/40/405502} {\bibfield  {journal}
  {\bibinfo  {journal} {Journal of Physics A: Mathematical and Theoretical}\
  }\textbf {\bibinfo {volume} {44}},\ \bibinfo {pages} {405502} (\bibinfo
  {year} {2011})}\BibitemShut {NoStop}%
\bibitem [{\citenamefont {Barr\'{e}}\ \emph {et~al.}(2010)\citenamefont
  {Barr\'{e}}, \citenamefont {Olivetti},\ and\ \citenamefont
  {Yamaguchi}}]{BarreOlivettiYamaguchi:jstat2010}%
  \BibitemOpen
  \bibfield  {author} {\bibinfo {author} {\bibfnamefont {J.}~\bibnamefont
  {Barr\'{e}}}, \bibinfo {author} {\bibfnamefont {A.}~\bibnamefont {Olivetti}},
  \ and\ \bibinfo {author} {\bibfnamefont {Y.~Y.}\ \bibnamefont {Yamaguchi}},\
  }\href {\doibase 10.1088/1742-5468/2010/08/P08002} {\bibfield  {journal}
  {\bibinfo  {journal} {Journal of Statistical Mechanics: Theory and
  Experiment}\ }\textbf {\bibinfo {volume} {2010}},\ \bibinfo {pages} {P08002}
  (\bibinfo {year} {2010})}\BibitemShut {NoStop}%
\bibitem [{\citenamefont {Chavanis}\ and\ \citenamefont
  {Campa}(2010)}]{ChavanisCampa}%
  \BibitemOpen
  \bibfield  {author} {\bibinfo {author} {\bibfnamefont {P.}~\bibnamefont
  {Chavanis}}\ and\ \bibinfo {author} {\bibfnamefont {A.}~\bibnamefont
  {Campa}},\ }\href {\doibase https://doi.org/10.1140/epjb/e2010-00243-x}
  {\bibfield  {journal} {\bibinfo  {journal} {The European Physical Journal B}\
  }\textbf {\bibinfo {volume} {76}},\ \bibinfo {pages} {581} (\bibinfo {year}
  {2010})}\BibitemShut {NoStop}%
\bibitem [{\citenamefont {Messer}\ and\ \citenamefont
  {Spohn}(1982)}]{MesserSpohn}%
  \BibitemOpen
  \bibfield  {author} {\bibinfo {author} {\bibfnamefont {J.}~\bibnamefont
  {Messer}}\ and\ \bibinfo {author} {\bibfnamefont {H.}~\bibnamefont {Spohn}},\
  }\href {\doibase https://doi.org/10.1007/BF01342187} {\bibfield  {journal}
  {\bibinfo  {journal} {Journal of Statistical Physics volume}\ }\textbf
  {\bibinfo {volume} {29}},\ \bibinfo {pages} {561} (\bibinfo {year}
  {1982})}\BibitemShut {NoStop}%
\bibitem [{\citenamefont {Antoni}\ and\ \citenamefont
  {Ruffo}(1995)}]{RUFFO1995}%
  \BibitemOpen
  \bibfield  {author} {\bibinfo {author} {\bibfnamefont {M.}~\bibnamefont
  {Antoni}}\ and\ \bibinfo {author} {\bibfnamefont {S.}~\bibnamefont {Ruffo}},\
  }\href {\doibase 10.1103/PhysRevE.52.2361} {\bibfield  {journal} {\bibinfo
  {journal} {Phys. Rev. E}\ }\textbf {\bibinfo {volume} {52}},\ \bibinfo
  {pages} {2361} (\bibinfo {year} {1995})}\BibitemShut {NoStop}%
\bibitem [{\citenamefont {Press}\ \emph {et~al.}(2007)\citenamefont {Press},
  \citenamefont {Teukolsky}, \citenamefont {Vetterling},\ and\ \citenamefont
  {Flannery}}]{press2007numerical}%
  \BibitemOpen
  \bibfield  {author} {\bibinfo {author} {\bibfnamefont {W.}~\bibnamefont
  {Press}}, \bibinfo {author} {\bibfnamefont {S.}~\bibnamefont {Teukolsky}},
  \bibinfo {author} {\bibfnamefont {W.}~\bibnamefont {Vetterling}}, \ and\
  \bibinfo {author} {\bibfnamefont {B.}~\bibnamefont {Flannery}},\ }\href@noop
  {} {\emph {\bibinfo {title} {Numerical Recipes: The art of Scientific
  Computing, Third Edition in C++}}}\ (\bibinfo  {publisher} {Cambridge
  University Press},\ \bibinfo {year} {2007})\BibitemShut {NoStop}%
\bibitem [{\citenamefont {{de Buyl}}(2014)}]{DEBUYL20141822}%
  \BibitemOpen
  \bibfield  {author} {\bibinfo {author} {\bibfnamefont {P.}~\bibnamefont {{de
  Buyl}}},\ }\href {\doibase https://doi.org/10.1016/j.cpc.2014.03.004}
  {\bibfield  {journal} {\bibinfo  {journal} {Computer Physics Communications}\
  }\textbf {\bibinfo {volume} {185}},\ \bibinfo {pages} {1822 } (\bibinfo
  {year} {2014})}\BibitemShut {NoStop}%
\bibitem [{\citenamefont {{Cheng}}\ and\ \citenamefont
  {{Knorr}}(1976)}]{1976JChengKnorr}%
  \BibitemOpen
  \bibfield  {author} {\bibinfo {author} {\bibfnamefont {C.~Z.}\ \bibnamefont
  {{Cheng}}}\ and\ \bibinfo {author} {\bibfnamefont {G.}~\bibnamefont
  {{Knorr}}},\ }\href {\doibase 10.1016/0021-9991(76)90053-X} {\bibfield
  {journal} {\bibinfo  {journal} {Journal of Computational Physics}\ }\textbf
  {\bibinfo {volume} {22}},\ \bibinfo {pages} {330} (\bibinfo {year}
  {1976})}\BibitemShut {NoStop}%
\bibitem [{\citenamefont {Sonnendr{\"u}cker}\ \emph {et~al.}(1999)\citenamefont
  {Sonnendr{\"u}cker}, \citenamefont {Roche}, \citenamefont {Bertrand},\ and\
  \citenamefont {Ghizzo}}]{SONNENDRUCKER1999201}%
  \BibitemOpen
  \bibfield  {author} {\bibinfo {author} {\bibfnamefont {E.}~\bibnamefont
  {Sonnendr{\"u}cker}}, \bibinfo {author} {\bibfnamefont {J.}~\bibnamefont
  {Roche}}, \bibinfo {author} {\bibfnamefont {P.}~\bibnamefont {Bertrand}}, \
  and\ \bibinfo {author} {\bibfnamefont {A.}~\bibnamefont {Ghizzo}},\ }\href
  {\doibase https://doi.org/10.1006/jcph.1998.6148} {\bibfield  {journal}
  {\bibinfo  {journal} {Journal of Computational Physics}\ }\textbf {\bibinfo
  {volume} {149}},\ \bibinfo {pages} {201 } (\bibinfo {year}
  {1999})}\BibitemShut {NoStop}%
\bibitem [{\citenamefont {Antoniazzi}\ \emph {et~al.}(2007)\citenamefont
  {Antoniazzi}, \citenamefont {Fanelli}, \citenamefont {Barr\'e}, \citenamefont
  {Chavanis}, \citenamefont {Dauxois},\ and\ \citenamefont
  {Ruffo}}]{Antoniazzi2007}%
  \BibitemOpen
  \bibfield  {author} {\bibinfo {author} {\bibfnamefont {A.}~\bibnamefont
  {Antoniazzi}}, \bibinfo {author} {\bibfnamefont {D.}~\bibnamefont {Fanelli}},
  \bibinfo {author} {\bibfnamefont {J.}~\bibnamefont {Barr\'e}}, \bibinfo
  {author} {\bibfnamefont {P.-H.}\ \bibnamefont {Chavanis}}, \bibinfo {author}
  {\bibfnamefont {T.}~\bibnamefont {Dauxois}}, \ and\ \bibinfo {author}
  {\bibfnamefont {S.}~\bibnamefont {Ruffo}},\ }\href {\doibase
  10.1103/PhysRevE.75.011112} {\bibfield  {journal} {\bibinfo  {journal} {Phys.
  Rev. E}\ }\textbf {\bibinfo {volume} {75}},\ \bibinfo {pages} {011112}
  (\bibinfo {year} {2007})}\BibitemShut {NoStop}%
\end{thebibliography}%

\end{document}